%% file: 20240521_main.tex
\RequirePackage{plautopatch}
\documentclass[a4paper,11pt]{article}

\usepackage{ulem} 

\usepackage{geometry}
\usepackage{here}

\usepackage{multicol}
\usepackage{multirow}
\usepackage{fancybox, ascmac}

\usepackage[authoryear]{natbib}
\usepackage[hyphens]{url}
\usepackage{hhline}

\usepackage{amsmath, amssymb} 
\usepackage{bm} 
\usepackage{array, booktabs} 
\usepackage[utf8]{inputenc}
\usepackage{txfonts}
\usepackage[T1]{fontenc}

\usepackage{palatino}

\usepackage[pdftex]{graphicx}
\usepackage{tikz}
\usepackage[framemethod=tikz]{mdframed}
\usepackage{hyperref}
\hypersetup{
    colorlinks = true,
    linkcolor  = blue,
    citecolor  = blue,
    filecolor  = purple,      
    urlcolor   = blue,
    }

\usepackage{tabularx} 
\usepackage{colortbl} 
\usepackage{float} 

\def\diamondleaders{\par\vskip.5\baselineskip
 \leavevmode\hbox{}\hskip1zw\leaders
 \hbox to.5zw{\hss\footnotesize ◇\hss}
 \hfill\hskip1zw\hbox{}\par}

\makeatletter
\def\thanks#1{%
   \footnotemark
   \edef\@tempa{\noexpand\noexpand\noexpand\footnotetext[\the\c@footnote]}%
   \toks@\expandafter{\@thanks}%
   \toks\tw@{{#1}}
   \xdef\@thanks{\the\toks@\@tempa\the\toks\tw@}}
\makeatother

\begin{document}

\author{Daisuke Ida\thanks{Faculty of Economics, Momoyama Gakuin University, 
                           1-1, Manabino, Izumi, Osaka 594-1198, Japan.
 			   Tel.: +81-725-54-3131.
 			   E-mail: ida-dai@andrew.ac.jp} and 
    Hirokuni Iiboshi\thanks{Faculty of Economics, Nihon University.
			   E-mail: iiboshi.hirokuni@nihon-u.ac.jp}}
\date{\today}
\title{The international forward guidance transmission \\ under a global liquidity trap\thanks{The previous title of this paper was ``The interaction of forward guidance in a two-country new Keynesian model."  We would like to thank Takashi Kano, Nao Sudo, and the conference participants at the autumn meeting of JEA 2020 for their valuable comments and suggestions. Ida was supported by JSPS KAKENHI Grant Numbers 16H03618, 20K01784, and JP24K04971. Iiboshi was supported by JSPS KAKENHI Grant Numbers 18K01575 and 19H01494.}}

\maketitle

\begin{abstract}
This paper quantitatively explores the interaction effect of forward guidance (FG) on international monetary policy transmission using a standard two-country new Keynesian model with a global liquidity trap. First, we show that the magnitude of the constant risk aversion coefficient (CRRA) is important in determining the beggar-thy-neighbor and prosper-thy-neighbor effects in foreign economies when the home country only faces the zero lower bound (ZLB) constraints. Second, we demonstrate that both countries may benefit from adopting only the home country's FG policies if the home central bank only faces the ZLB. Third, we find the potential benefit of the FG interaction effect between two countries. Thus, we document the possibility that home and foreign central banks can benefit from monetary policy coordination by adopting the same duration of FG quarters.
\end{abstract}
\begin{verse}
JEL codes: E52; E58; F41 \\
Keywords: Forward guidance; Taylor rule; Zero lower bound on nominal interest rates; Two-country new Keynesian model; \\
\end{verse}

\thispagestyle{empty}
\pagebreak
\setcounter{page}{1}

\section{Introduction}
As long as nominal interest rates remain above zero, the central bank can manipulate nominal interest rates as a primary policy instrument of conventional monetary policy. However, once the policy rate reaches the zero lower bound (ZLB), the central bank loses the ability to control the short-term nominal interest rate. As a result, the ZLB constraint prevents further policy rate reductions. Accordingly, over the last decade, theoretical and empirical research has focused on a variety of unconventional monetary policies. Among them, forward guidance (FG) is probably the most notable monetary policy tool under the ZLB \citep{bernanke2004conducting,cole_2023, haberis2019uncertain,haberis2020welfare,jung2005optimal, moessner2019zero,woodford2003zero}. FG indicates that, in the presence of a ZLB constraint, the central bank can stimulate the economy by promising to keep future interest rates at zero for a while, even if future increases in aggregate demand suggest the termination of the zero interest rate policy (ZIRP). Indeed, as shown in Figure 1, central banks in advanced economies used zero interest rate policies as an unconventional monetary policy tool to combat deflationary pressures resulting from the global financial crisis. Central banks have recognized the issue of conducting monetary policy under the ZLB as a global challenge. As a result, many central banks adopted the FG policy as a standard policy tool while adhering to ZLB constraints.


\begin{figure}[h]
\caption{Policy rates in advanced countries}
\begin{center}
\includegraphics[width=10cm,height=10cm]{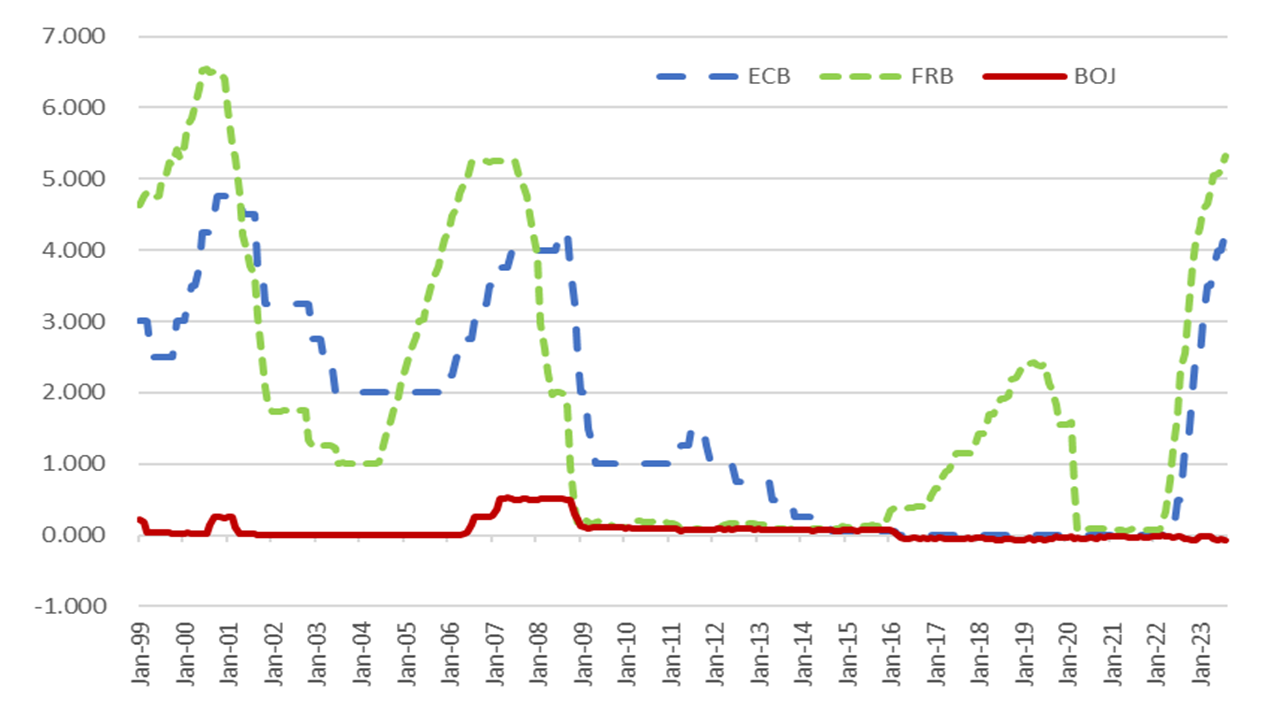}

 \begin{footnotesize}
 Source: Federal Reserve Bank of St.Louis. Percentage points. 
\end{footnotesize}
\end{center}
\end{figure}

The global aspect of FG documents how the termination of one country's FG spills over the other countries through a change in exchange rates. For example, the taper tantrum shock of the Federal Reserve Chairman Bernanke's testimony to Congress on May 22, 2013, triggered a global drop in stock prices. This shock could be caused by the fact that the stock market participants expected the central bank to reconsider the FG strategy. Recently, inflationary pressures in the US economy compelled the Federal Reserve Bank (FRB) to aggressively raise the Federal Funds rate. The FRB's aggressive monetary contraction has caused the Japanese yen to fall sharply against the US dollar. As shown in Figure 1, this was primarily due to the substantial interest rate differential between Japan and the US caused by the sharp increase in US policy rates. These facts suggest that each central bank's FG termination date is important in accounting for international monetary policy transmission. However, to the best of our knowledge, only a few studies have attempted to examine the interaction effect of home and foreign FG from a global economic perspective. Assessing the interaction effect of FG policies in an open economy context is extremely important.

To fill this gap, we consider the importance of an international FG transmission mechanism in a two-country model with a global liquidity trap. The first contribution is to explore how and to what extent the unconventional monetary policies implemented by countries under the zero interest rate regime after the global financial crisis affect each other and how this interaction influences policy decisions between countries. To accomplish this, this study uses the standard two-country new Keynesian (NK) model developed by \citet{clarida2002simple}. We extend this widely used two-country model to account for the situation in which two countries face ZLB constraints simultaneously. The use of the NK model to study unconventional monetary policy in closed economies under the ZLB is becoming more common. This is due to the development of numerical methods and the performance of the hardware that enables them. However, there appears to be a limited number of papers that discuss how FG in one country spills over to other countries and the international transmission mechanism of monetary policy \citep{haberis2020welfare,jones2018international}.\footnote{\citet{ida2023effect} examined the effect of a money growth rule, which becomes a proxy as quantitative easing, on international monetary transmission mechanism in a two-country NK model with the ZLB.}

When both countries face the ZLB constraint, we use this two-country NK model to analyze the international spillover effects of implementing FG policy. We evaluate policies in two scenarios in which two countries face the ZLB simultaneously: a) when one of the countries' central banks adopts FG policies; and b) when both central banks adopt FG policies simultaneously. We then examine how extending the duration of each country's FG policy affects output, inflation, and nominal interest rates in both countries through international monetary policy transmission. Along with the international spillover effects of FG, we also compare the welfare losses in the two countries. The numerical simulations show what kind of FG interaction the two countries can expect. 

The numerical method that we employ is developed by \citet{guerrieri2015occbin} and is referred to as occasionally binding constraints (OccBin) toolkit. The second contribution is to use the OccBin toolkit to capture international FG transmission during a liquidity trap. Numerous studies, including \citet{guerrieri2015occbin}, use it for unconventional monetary policy analysis under the ZLB in a closed-economy model. A study of the application of the OccBin toolkit to FG policy in the two-country NK model was conducted by \citet{haberis2020welfare}. \citet{ida2023effect} also used the OccBin toolkit to think about the role of money growth targeting in a two-country model with ZLB constraints. To the best of our knowledge, our study appears to be the first to use the toolkit to simultaneously implement FG policies in a two-country NK model.

The findings of this paper are summarized as follows. First, the magnitude of the constant risk aversion coefficient (CRRA) is important in determining the beggar-thy-neighbor and prosper-thy-neighbor effects in foreign economies when the home country only faces the ZLB. In other words, when the CRRA coefficient is less than one, home FG has a negative economic effect on the foreign country with a symmetrical economic structure. If the CRRA coefficient is greater than unity, home FG has the inverse economic effect on the counterpart country. Second, both countries may benefit from adopting only the home country's FG policies when the home central bank only faces the ZLB. Furthermore, extending the duration of FG over its counterpart will result in a duration of FG that minimizes the home country's welfare loss. Third, we discuss the potential benefit of the FG interaction effect between two countries. Thus, we argue the possibility that by adopting the same duration of FG quarters, home and foreign central banks can benefit from monetary policy coordination. It should be noted, however, that such international policy cooperation may not always be possible because home and foreign central banks have incentives to deviate from such cooperation, and further investigation of this issue is beyond the scope of this paper.

This paper is organized as follows. Section 2 reviews the literature relevant to this study and Section 3 describes the two-country NK model. Section 4 presents the main findings of this paper. We examine the interaction effect of home and foreign FG policies in an open economy and provide some robustness checks. Section 5 discusses policy implications based on the results of our paper. Section 6 provides a brief conclusion to the paper.

\section{Related literature and our contribution}
In this section, we provide a brief review of previous literature on the effects of unconventional monetary policy using the NK model. First, we categorize the effects of the FG policy in a closed-economy NK model. Second, we review the role of unconventional monetary policy in an open economy. Third, we describe how our findings serve as an extension of those in previous studies. 

\subsection*{\textit{The FG effectiveness}}
As previously stated, the ZLB has recently become a major issue in advanced economies such as the United States, the Eurozone, and Japan. Moreover, once faced with the ZLB, central banks cannot manipulate short-term nominal interest rates as a policy tool.\footnote{See \citet{bernanke2004conducting}, \citet{englishfederal}, and \citet{williamson2015monetary} for a detailed discussion of unconventional monetary policy.} Within this framework, a number of studies have addressed the expectations channel of monetary policy by managing private-sector expectations. In particular, the NK-model-based theoretical studies strongly support the effectiveness of this expectations channel. For example, \citet{woodford2003zero} and \citet{jung2005optimal} argue that even in the face of liquidity traps, central banks can exert monetary policy effects. According to their study, even in the case of a liquidity trap caused by a negative natural interest rate, the central bank can manipulate private-sector expectations by committing to keep interest rates at zero for an extended period.\footnote{ According to \citet{adam2006optimal} and \citet{adam2007discretionary}, the ZLB increases the cost of implementing discretionary policy compared to a non-ZLB model. \citet{hirose2023estimating} examine the effect of FG of behavioral NK model for the US economy, whereas \citet{iiboshi2022estimating} estimate the impact of FG policy on the Japanese's economy.}  

The FG policy's key mechanism relys on the strength of history dependence via lagged endogenous variables in the forward-looking NK model. However, as argued in \citet{del2012forward}, \citet{carlstrom2015inflation}, and \citet{mckay2016power}, there is little evidence that the FG policy was successful in creating a substantial increase in inflation and the output gap; therefore, this paradoxical phenomenon is known as the FG puzzle. To address this issue in an NK model, several studies departed from the assumption of rational expectations \citep{gabaix2020behavioral}.

\subsection*{\textit{Unconventional monetary policy in an open economy}}
Several studies have also examined the effectiveness of unconventional monetary policy in an open economy. First, consider the role of the FG policy in an open economy. \citet{gali2020uncovered} showed that the difference between home and foreign interest rates captures the effect of FG policy in a small open economy model.\footnote{He argues that the exchange rate dynamics are incompatible with the interest rate parity hypothesis. He calls this inconsistency the forward guidance exchange rate puzzle.} \citet{haberis2020welfare} also examined the international transmission mechanism of FG policy in a small open economy model. According to their analysis, foreign FG policies are effective when they improve social welfare in the home country. \citet{jones2018international} estimated the international spillover effects of FG policies in a two-country NK model. They showed that when monetary easing FG shocks occur in the US economy, they have a beggar-thy-neighbor effect on the Canadian economy. Thus, when the US economy experiences a monetary easing FG shock, the Canadian economy's output will suffer significantly.

Second, we review previous studies that have focused on policy interactions between two countries. As previously stated, central banks in advanced economies faced the ZLB following the 2008 global financial crisis and should have considered such a financial crisis to be the source of the negative global demand shock. Therefore, several papers have discussed the topic of ZLBs in open economies \citep{cook2011optimal,fujiwara2013global,ida2013optimal,nakajima2008liquidity}. \citet{fujiwara2013global} examined the optimal commitment policy in a two-country NK model where the home country and the foreign country face a ZLB. Their analysis shows that as negative natural interest rate shocks in each country return to a steady state, each country's optimal commitment policy becomes more complex than in a closed-economy model. They also show the effectiveness of price level targeting in a two-country NK model where two countries face a ZLB. Meanwhile, \citet{alpanda2020international} and \citet{kolasa2020international} attempted to consider the international spillover path of quantitative easing (QE) in terms of the central bank's asset purchases. \citet{ida2023effect} explored the effect of QE via an increase in money supply on international monetary transmission mechanisms and showed that when the home and foreign countries face the ZLB, a foreign QE shock may have a beggar-thy-neighbor effect on the home country.

\subsection*{\textit{The contribution of this paper to previous studies.}}
The contributions of this study are summarized as follows. First, this study is closely related to \citet{haberis2020welfare}, who examined the impact of large foreign FG policies on a small home country. In their analysis, the large country's central bank follows the monetary policy rules imposed by the FG, whereas the small country's central bank can implement its optimal commitment policy. In contrast, our analysis assumes that both the home and foreign central banks of large countries implement monetary policy in accordance with the Taylor rule with the FG under the ZLB. It also focuses on the interaction effect of the home and foreign central banks' FG. This study concentrates on the more practical aspects of monetary policy rather than the optimal monetary policy. Moreover, although \citet{haberis2020welfare} consider that negative shocks to the natural rate of interest occur only in foreign countries, our study addresses the case of a global liquidity trap, in which natural interest rates fall simultaneously in the home country and abroad. 

Second, this study shows that the magnitude of the risk aversion coefficient of the CRRA coefficient in the utility function is important in determining the beggar-thy-neighbor and prosper-thy-neighbor effects in foreign economies. In a two-country model without the ZLB, \citet{tille2001role} explored the relationship between consumption substitutability and international monetary policy spillover. \citet{tille2001role} focused on the relationship between consumption substitutability and international spillover of monetary policy in the absence of the ZLB. While we consider this relationship, we also note that whether monetary easing in one country has beggar-thy-neighbor or prosper-thy-neighbor effects on the other is unclear in the context of a global liquidity trap.\footnote{While this study investigates an international spillover effect of a FG policy in a two-country model developed by \citet{clarida2002simple}, \citet{tille2001role} consider the effect of consumption substitutability on optimal monetary policy in \citet{obstfeld1996foundations}'s model. } 

Third, our paper is also close to that of \citet{fujiwara2013global}, who examine the interaction of ZLB policies between central banks in a two-country NK model with a global liquidity trap regarding optimal commitment policy. In contrast to their study, our study considers the case in which home and foreign central banks use the simple Taylor rule with FG instead of implementing optimal commitment policy. In comparison to the closed economy, it is understandable that central banks find it difficult to implement an optimal commitment solution under policy coordination. Therefore, as mentioned earlier, the more practical aspects of monetary policy justify the assumption of using a simple monetary policy rule, the more the use of a simple Taylor rule allows us to examine how the different lengths of each country's FG affect international monetary policy transmission during a liquidity trap. Thus, applying a simple monetary policy rule can help clarify the policy implications suggested in this paper.

Fourth, our paper also contributes to previous studies that investigated the effects of unconventional monetary policy in an open economy. \citet{kolasa2020international} focused on the role of asset purchases as a QE policy tool and argued that the role of the large country's asset purchases on the small country is mediated by a change in the term premium. \citet{ida2023effect} addressed an important aspect of QE policy by demonstrating how an increase in the money supply boosts output and inflation in a two-country model with the ZLB. Meanwhile, \citet{wu2019global} examined the role of unconventional monetary policy with a shadow rate of the policy rate in a two-country NK model. However, none of these studies examined the interaction effect of FG policies in a two-country NK model.

\section{A simple two-country NK model}
This study adopts the simple two-country NK framework developed by \citet{clarida2002simple} to intuitively capture the international spillover effects of forward guidance.\footnote{This study considers a two-country model in which the consumption basket consists of Cobb-Douglus aggregates. \citet{pappa2004ecb} considered the costs without policy coordination in a two-country NK model where the consumption basket consists of constant elasticity of substitution aggregates.} Consider an economy with two large, symmetrical countries: home and foreign. The population sizes of the home and foreign countries are $1-\gamma$ and $\gamma$, respectively. Each country has two production sectors: the final goods sector, which is characterized by perfect competition, and the intermediate goods sector, where firms face monopolistic competition and nominal price rigidity \citep{calvo1983staggered}. Assume both countries have a complete market and only final goods are traded. Each country has the same number of final goods producers as there are households. Because the model assumes purchasing power parity, we consider the case of producer currency pricing. 

Finally, unless otherwise noted, the same equation applies to foreign countries. Asterisks represent variables for foreign countries.

\subsection{A log-linearized model}
We briefly present a structural equation system that is log-linearized.\footnote{The online Appendix provides a detailed derivation of the structural model. See also \citet{clarida2002simple} and \citet{walsh2017monetary}.} The structural equations are log-linearized around the steady state. Here, lowercase variables are used to represent the logarithmic deviation from the steady state. Specifically, the log-linearized variables around the steady state are represented by $h_t=\log(H_t/\bar{H})$, where $\bar{H}$ represents the steady state value. The log-linearized structural equations can be summarized as follows:
\begin{align}
&\pi_t = \beta E_t \pi_{t+1} + \kappa_1 x_t + \kappa_2 x^*_t, \label{eq.hnkpc1}
\\
&\pi^*_t = \beta E_t \pi^*_{t+1} + \kappa_1^* x^*_t + \kappa_2^* x_t, \label{eq.fnkpc2}
\\
&x_t = E_t x_{t+1} +\vartheta E_t\Delta x^*_{t+1} -(\sigma_0)^{-1}(r_t - E_t \pi_{t+1}  - r_t^n), \label{eq.his3}
\\
&x^*_t = E_t x^*_{t+1} +\vartheta^* E_t\Delta x_{t+1} -(\sigma_0^*)^{-1}(r^*_t - E_t \pi^*_{t+1}  - (r_t^n)^*), \label{eq.fis4}
\\
&r_t \ge 0, \label{eq.hzlb}
\\
& r_t^* \ge 0. \label{eq.fzlb}
\end{align}
$\pi_t$ is the home country inflation, $\pi^*_t$ is the foreign country inflation, $x_t$ is the home country output gap, and $x^*_t$ is the foreign country output gap. $r_t$ is the home country's nominal interest rate and $r^*_t$ is the foreign country's interest rate. $r^n_t$ is the home country natural interest rate and $(r^n_t)^*$ represents the foreign natural interest rate. The shock to each country's natural rate of interest follows an AR(1) process. The next subsection describes how the natural rate of interest evolves in a two-country model. Additionally, the structural parameters are defined as follows:
\begin{align}
&\kappa_1 = \frac{(1-\alpha)(1-\alpha\beta)}{\alpha}(\sigma+\eta-\gamma(\sigma-1)); \ \kappa_2 = \frac{(1-\alpha)(1-\alpha\beta)}{\alpha}\gamma(\sigma-1); \nonumber \\
&\kappa_1^* = \frac{(1-\alpha^*)(1-\alpha^*\beta)}{\alpha^*}(\sigma+\eta-(1-\gamma)(\sigma-1)); \ \kappa_2^* = \frac{(1-\alpha^*)(1-\alpha^*\beta)}{\alpha^*}(1-\gamma)(\sigma-1); \nonumber \\
&\vartheta = \frac{\gamma(\sigma-1)}{\sigma-\gamma(\sigma-1)};  \ \vartheta^* = \frac{(1-\gamma)(\sigma-1)}{\sigma-(1-\gamma)(\sigma-1)}; \ \sigma_0 = \sigma -\gamma(\sigma-1); \ \sigma_0^* = \sigma -(1-\gamma)(\sigma-1). \nonumber
\end{align}
$\sigma$ and $\eta$ denote the constant relative risk aversion coefficient and the inverse elasticity of labor supply. As previously noted, the parameter $\gamma$ represents the degree of openness. $\beta$ and $\alpha$ denote the discount factor and the degree of nominal price stickiness, namely, the Calvo lottery. Equations (\ref{eq.hnkpc1}) and (\ref{eq.fnkpc2}) represent open economy new Keynesian Phillips curve (NKPC).\footnote{See \citet{clarida2002simple} for a detailed discussion of these channels.} Equations (\ref{eq.his3}) and (\ref{eq.fis4}) represent the open economy dynamic IS (DIS) curve. Finally, (\ref{eq.hzlb}) and (\ref{eq.fzlb}) denote the ZLB constraints on nominal interest rates for each country.

The foreign output gap influences both the DIS curve and the NKPC in the home country, via risk sharing and terms of trade. The term $\gamma(\sigma-1)$ includes the risk sharing and the terms of trade effects. We label these channels as the international spillover effects. As explained by \citet{clarida2002simple}, the risk sharing and terms of trade effects are captured by $\gamma\sigma$ and $\gamma$, respectively. On the one hand, the risk-sharing effect can be regarded as a change in consumption through the terms of trade in the real marginal cost. On the other hand, a change in the terms of trade directly affects the real marginal cost. The former effect dominates the latter one when $\sigma>1$, whereas the opposite case occurs in the case of $\sigma<1$. These effects are the cause of the foreign output gap in the home structural equations. A similar discussion applies to the foreign economy. Notice that international spillover effects vanish when the parameter $\sigma$ becomes unity.

\subsection{Monetary policy rules, FG specification, and structural shocks}
\subsubsection*{\textit{Monetary policy rules}}
We specify monetary policy rules. To consider the role of the FG policy, we adopt a simple instrument rule like the Taylor rule in contrast to \citet{fujiwara2013global} and \citet{haberis2020welfare}, who considered the role of a targeting rule associated with optimal monetary policy in an open economy. More precisely, following \citet{taylor1993discretion}, we assume that both home and foreign central banks adopt the following simple monetary policy rule, with the ZLB constraints on nominal interest rates:
\begin{align}
& r_t = \max\{0,\ (1-\psi_r)(\psi_{\pi} \pi_t+\psi_x x_t)+\psi_r r_{t-1} + e_t\}, \label{eq.hmprule} \\
& r^*_t = \max\{0, \ (1-\psi^*_r)(\psi^*_{\pi} \pi^*_t+\psi^*_x x^*_t)+\psi^*_r r^*_{t-1} + e^*_t\}. \label{eq.fmprule} 
\end{align}
Here, $\psi_{\pi}$ denotes the inflation stabilization, $\psi_x$ denotes the output gap stabilization, and $\psi_r$ is the term for interest rate smoothing. Additionally, $e_t$ represents an exogenous monetary policy shock, which assumes no shock persistence.

\subsubsection*{\textit{Specification of FG policies}}
Let us explain how FG is specified in this paper. Following \citet{haberis2020welfare}, the calendar-based FG policies adopted by both countries imply that the central bank commits to a ZIRP for longer than the ZLB periods suggested by the standard Taylor rule. In this study, we regard the terminology of FG policy adopted in \citet{haberis2020welfare} as the terminology of fixed-length FG suggested by \citet{eggertsson2020toolkit}. Following \citet{haberis2020welfare}, we introduce several specifications for FG strategies. Our simulation focuses on four types of FG specifications: two extra quarters, four extra quarters, five extra quarters, and ten extra quarters. In addition, we consider four policy options for the foreign central bank. First, the foreign central bank can carry out monetary policy in an economy without the ZLB. Second, like the home country, the foreign country faces the ZLB. Third, the foreign central bank follows the same length of FG policy as the home central bank. Fourth, while the home central bank can select any length from four types of FG policies, the foreign central bank sets the number of quarters of the FG policy to five. 

As mentioned earlier, \citet{haberis2020welfare} explored the international FG spillover effect of how a large country's FG policy affects an optimal commitment policy adopted by a small country. Unlike their study, we address how a global liquidity trap shock affects both large countries. This is because a negative shock to one country's natural interest rate exacerbates the situation in the other. As a result, we consider a global liquidity trap shock that has a negative impact on the natural rates of interest in both countries. Finally, we postulate that this shock causes an equal-sized decline in each country's natural interest rate. 

\subsubsection*{\textit{Structural shocks}}
This study considers two structural shocks: a shock to the natural rate of interest in only the home country and a global liquidity trap shock in both the home and foreign countries. Although the latter shock is distinguished by the former shocks, the latter feature is expected to occur simultaneously in both countries. These two shocks are assumed to be persistent, and they can be expressed as first-order autoregressive (AR) processes as follows:
\begin{align}
r_{t}^{n}&=\rho_{r}r_{t-1}^{n}+e_{t}^{NR}+e_{t}^{GL}, \nonumber \\
(r_{t}^{n})^{*}&=\rho_{r}(r_{t-1}^{n})^{*}+(e_{t}^{GL})^{*}, \nonumber
\end{align}
where $\rho_{r}$ is the coefficient of AR processes, $e_{t}^{NR}$ is the independent
shock to the natural rate of interest generated in the home country,
and $e_{t}^{GL}$ and$(e_{t}^{GL})^{*}$ are the independent shocks
to the global liquidity traps in the home and foreign countries, respectively.

\subsection{Welfare criteria}
To ensure consistency with this two-country NK model, we use simple welfare criteria to assess welfare losses associated with FG policies. More concretely, following \citet{clarida2002simple}, we consider the following central bank's loss function:\footnote{See Appendix B for a detailed derivation of the loss function.}
\begin{align}
L^w_t = (1-\psi)L_t + \psi L^*_t-2\Lambda x_t x^*_t. \label{eq.lossfun}
\end{align}
When home and foreign central banks consider their monetary policy together, this loss function can be considered. This loss function is used to evaluate the global welfare loss for various FG specifications. Notably, it is also beneficial to consider that the loss function for the home country ($L_t$) and that for the foreign country ($L^*_t$) are given as follows:
\begin{align}
& L_t = \pi_t^2 + \lambda_x x_t^2 + \lambda_r r_t^2, \label{eq.hloss} \\
& L^*_t = (\pi_t^*)^2 + \lambda_x^* (x_t^*)^2 + \lambda_r^* (r_t^*)^2. \label{eq.floss}
\end{align}
Furthermore, the structural parameters in home and foreign central banks' loss functions are defined as follows:
\begin{align}
&1-\psi = \frac{(1-\gamma)\varpi^{-1}}{(1-\gamma)\varpi^{-1}+\gamma(\varpi^*)^{-1}};\ \lambda_x = \frac{\kappa_1}{\theta}; \ \lambda_r = \frac{\eta_r}{\bar{v}\theta}; \ \lambda^*_x = \frac{\kappa^*_1}{\theta}; \ \lambda^*_r = \frac{\eta_r^*}{\bar{v}^*\theta}; \nonumber \\
&\Lambda = \frac{2(1-\gamma)\gamma(1-\sigma)}{\varpi\theta}; \ \varpi = \frac{(1-\alpha)(1-\alpha \beta)}{\alpha}; \ \varpi^* = \frac{(1-\alpha^*)(1-\alpha^* \beta)}{\alpha^*}. \nonumber 
\end{align}
$\eta_r$ denotes the interest elasticity of money demand, $\bar{v}$ is the velocity of money demand, and $\theta$ is the elasticity of substitution of individual goods in each country. Equation (\ref{eq.hloss}) consists of three objectives. The first term on the right side represents the stabilization of home inflation. The second term on the right side represents the stabilization of the output gap. Finally, the third term on the right side reflects interest rate stabilization. As argued in \citet{woodford2003interest}, this term could be associated with the presence of ZLB constraints. The corresponding objectives apply to the foreign central bank's loss function. 

Note that in Equation (\ref{eq.lossfun}), following \citet{clarida2002simple}, we consider the presence of the third term on the right-hand side, which stems from the international spillover effect via both risk sharing and the terms of trade channels. Because this term disappears when $\sigma=1$, as in a closed-economy model, the above loss function is simply a weighted average of the home and foreign loss functions. It is worth noting that we do not assess the welfare loss in terms of monetary policy coordination. Thus, we do not pursue further investigation into the possibility that each central bank reneges on international policy coordination of FG policies.

\subsection{Parameterizations}
Following previous studies in the NK literature, this section describes the calibrated values used in this study. Table 1 summarizes the calibrated values for structural parameters. We set the Calvo lottery at 0.9. Although this parameter value could be slightly larger than that used in the standard NK literature, it is reported by \citet{boehl2024estimation}. Additionally, \citet{benigno2006jmcb} reported that several European economies have observed a higher level of nominal price stickiness, with a value of approximately 0.9. The discount factor is 0.985 and the degree of openness is set to 0.5. Thus, we consider the case in which the size of the home country equals that of the foreign country. 

\include{Table_1.tex}

We set the relative risk aversion coefficient for consumption to $2.0$ as a benchmark calibration. The selection of this value is based on the calibrated value in \citet{pappa2004ecb}. This value is crucial in a two-country NK model. \citet{clarida2002simple} showed that when this value is less than unity, the international risk-sharing effect is negative. Thus, if this value becomes unity, the model corresponds to an economy in which the open economy effect disappears. In particular, \citet{fujiwara2013global} and \citet{nakajima2008liquidity} argued the role of this parameter in a two-country NK model with the ZLB. Even in the case for simple monetary policy rules, the effect of FG policies in a two-country economy is affected by the value of the risk aversion coefficient. Hence, we describe several values of this parameter in the robustness check.

Finally, we select the parameters for monetary policy rules and the natural interest rate. We assume that both countries use the same parameterization for the monetary policy rule. Inflation stabilization $\psi_{\pi}$ is set to 1.25 and the output gap stabilization is set to 0.5. We set the term for interest rate smoothing to zero. When interest rate smoothing is considered, we conjecture that the reaction of the interest rate gradually changes even after an FG policy expires. The AR (1) coefficient for each country's natural interest rate is 0.8 with a variance of 1.0.

\section{Quantitative results}
This section explores the impact of FG policy on international monetary policy spillovers in light of the global liquidity trap. To do so, we isolate two phases as follows. The first scenario is the ``one-country-only liquidity trap" case, which we will discuss in Section 4.1. Section 4.2 discusses the second scenario, which involves a ``global liquidity trap" that occurs in two countries simultaneously. In these two cases, our analysis quantitatively examines the direction, duration, and magnitude of the global monetary policy transmission mechanism at work by comparing domestic and international liquidity traps, employing impulse response and welfare analysis. As previously mentioned, quantitative analysis, following the methodology of \citet{haberis2020welfare}, who analyzed the transmission of one country's FG to a foreign country, the OccBin toolkit on the Dynare platform developed by \citet{guerrieri2015occbin}, we use a piecewise linear perfect foresight algorithm to capture model dynamics under an occasional binding constraint.

\subsection{FG under a home liquidity trap}
As a first scenario, we examine a situation in which the home central bank faces ZLB constraints but the foreign central bank is not. This examination is useful for considering the practical evidence of when the Bank of Japan (BOJ) implemented a calendar-based FG strategy in February 1999 to alleviate deflationary pressures. At the same time, the FRB, unhindered by the ZLB, reduced its policy rate in response to deflationary risks. This historical context informs our analysis and provides important insights into the effects of international monetary policy transmission during this period. 

\pagebreak
\begin{figure}[]
\caption{The impulse response to a natural rate shock of a single country: High degree of CRRA ($\sigma = 2$) }
\includegraphics[width=14cm,height=14cm]{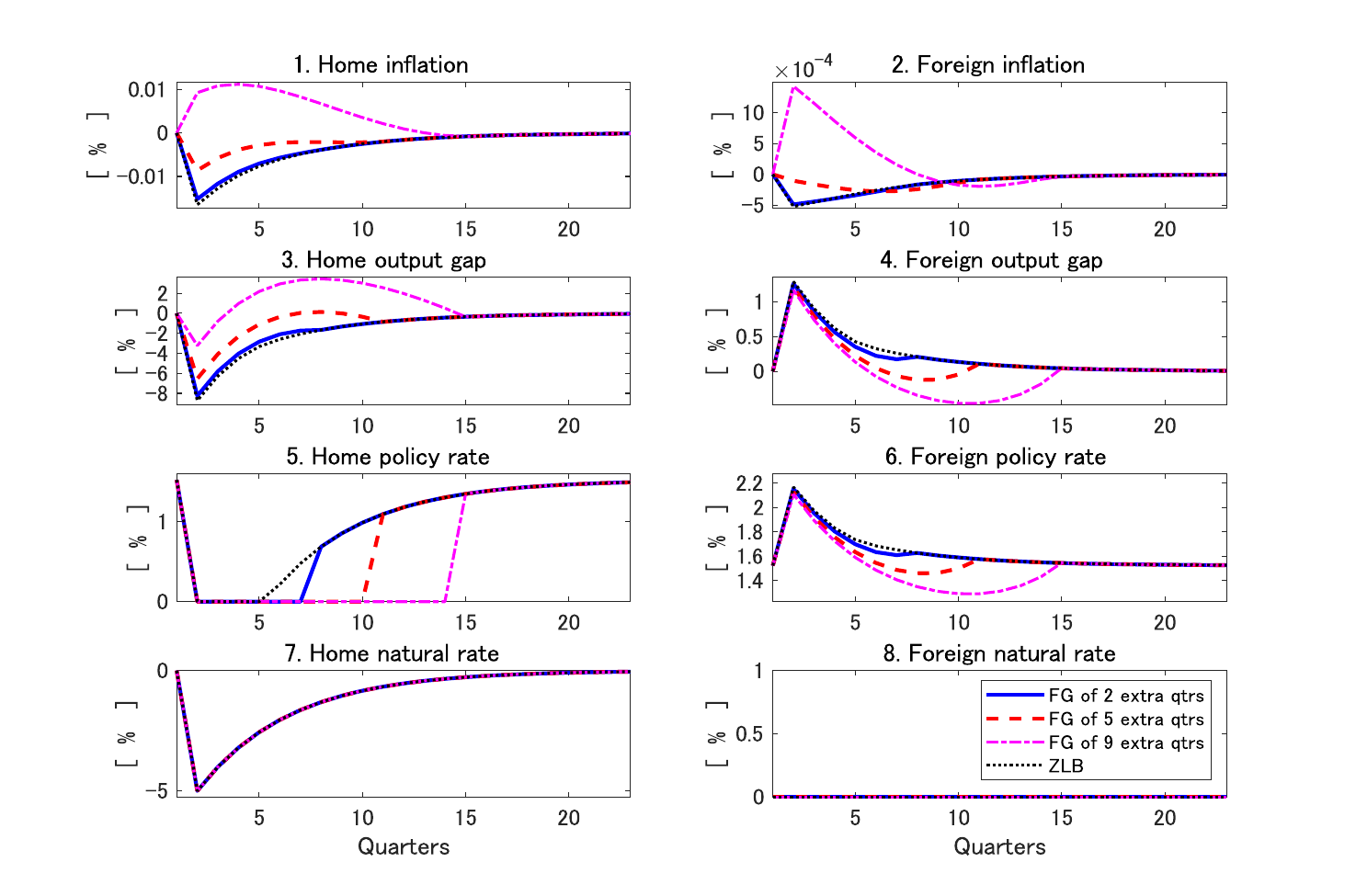}
\end{figure}

The first scenario in which the natural rate of interest declines only in the home country is the primary focus. Figure 2 depicts this case. The dotted line in Figure 2 shows that a shock to a negative natural interest rate reduces the home inflation rate and output gap. Following the shock, the nominal interest rate in the home country is subject to ZLB constraints. Conversely, when the domestic output gap declines, the foreign output gap increases. At period 5, the home central bank decides when the ZIRP will end, and nominal interest rates begin to rise gradually. If the home central bank implements a nine-period fixed-length FG, this situation will become substantial. In this case, the home central bank can completely alleviate the recession caused by a drop in the home natural rate of interest. The foreign country experiences a similar increase in inflation and the output gap, but it also experiences a severe recession following the boom.

\include{Table_2.tex}

The solid line represents the home country's two-period fixed-length FG. The graph shows that if the home central bank implements a shorter-term FG policy, it will be unable to stimulate inflation and the output gap. The dashed line depicts a five-period fixed-length FG for the home central bank. Unlike the previous two cases, this type of FG policy helps to mitigate the decline in the home country's inflation rate and output gap. In contrast, after a boom, foreign countries experience a significant decrease in their output gap.

How does the home central bank's FG adoption affect home and foreign welfare? Table 2(a) shows that, when compared to the case where the home central bank does not adopt FG, such FG adoption lowers both the home and foreign welfare losses. However, if the home central bank considers a longer-term FG, it increases both home and foreign welfare losses.

\begin{figure}[h]
\caption{The impulse response to a natural rate shock of a single country: High degree of CRRA ($ \sigma = 1$) }
\includegraphics[width=14cm,height=14cm]{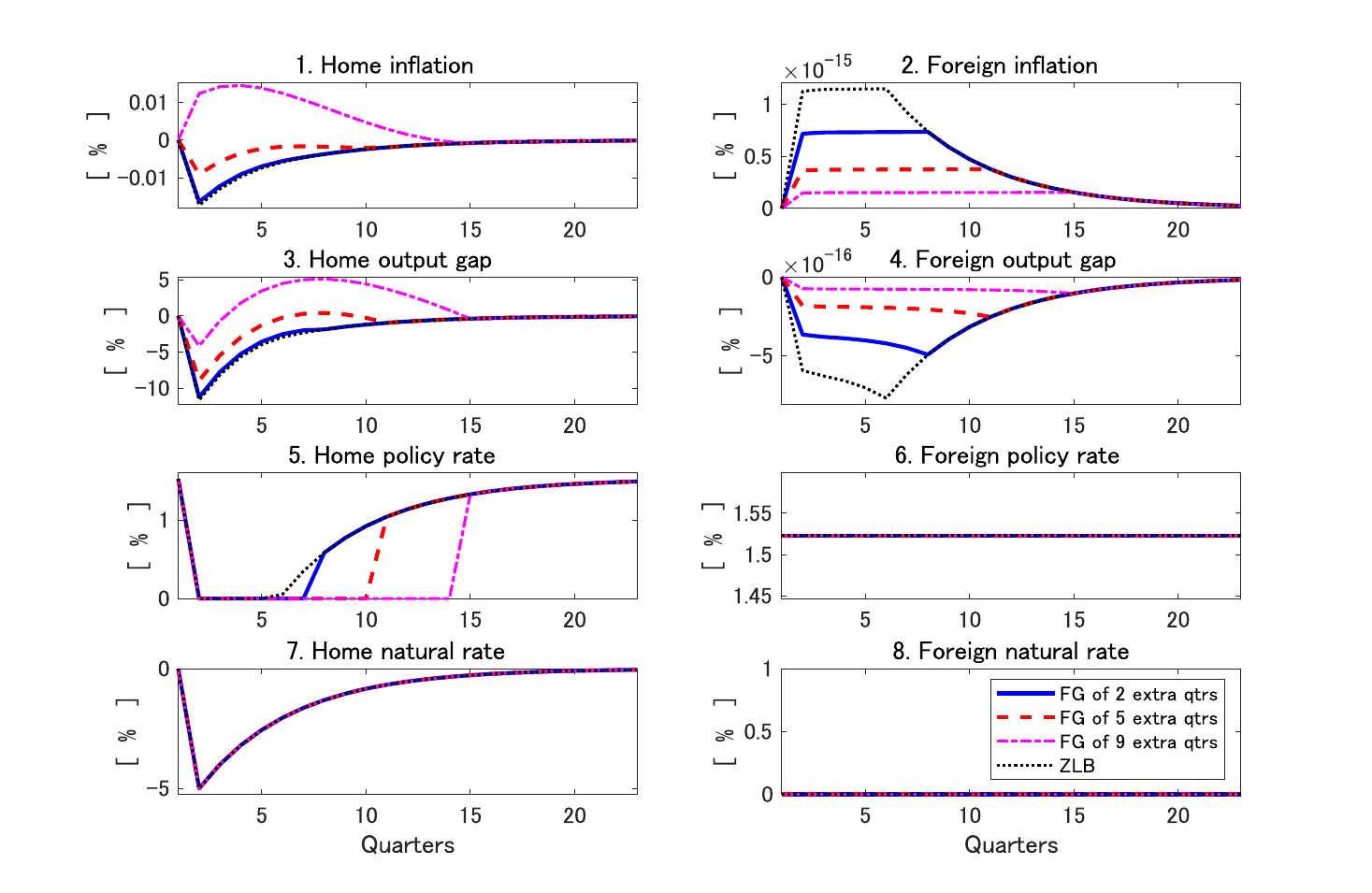}
\end{figure}

The above result is based on the case of $\sigma = 2.0$. As argued in \citet{clarida2002simple} and \citet{fujiwara2013global}, the value of a CRRA coefficient is critical for assessing international monetary policy transmission. Figure 3 depicts the effect of home FG on international monetary policy transmission under a home liquidity trap in the case of $\sigma = 1.0$. As shown in Figure 3 and Table 2(b), international monetary transmission and welfare losses remain unaffected when we set the parameter $\sigma$ to unity.

\begin{figure}[h]
\caption{The impulse response to a natural rate shock of a single country: High degree of CRRA ($ \sigma = 0.5$) }
\includegraphics[width=14cm,height=14cm]{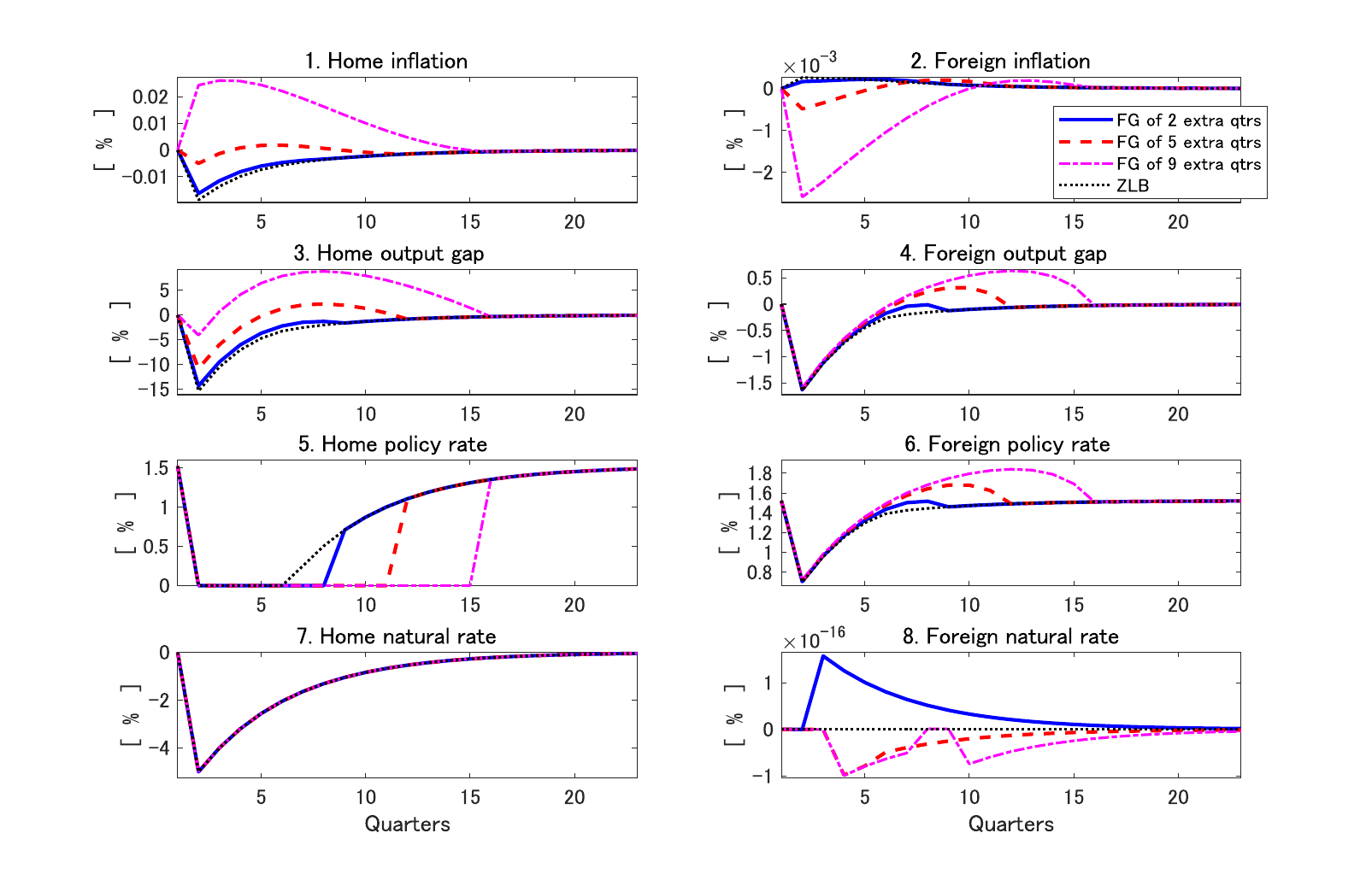}
\end{figure}


Figure 4 shows the case of $\sigma < 1$. In this case, according to the NKPC, when $\sigma < 1$, an increase in the foreign output gap will lower home inflation rates. In the home country, the central bank can stimulate the economy by reinforcing FG policy. Interestingly, a home liquidity trap shock causes a recession in the foreign country. As a result, the foreign central bank reduces its nominal interest rate because it is not subject to ZLB constraints. Particularly, as the home central bank employs a strengthened FG, the foreign output gap expands, decreasing the foreign inflation rate. Table 2(c) reports the welfare loss in the case of $\sigma < 1$. The results show that when the home central bank uses a five-period fixed-length FG, it incurs the least amount of welfare loss. However, when the four-period fixed-length FG is implemented in the home country, the foreign country experiences the least welfare loss. When the home central bank implements a five-period fixed-length FG, the global welfare loss is minimal.

In conclusion, if the home country only has a ZLB constraint, the home central bank can increase inflation and the output gap by implementing a long FG. The CRRA coefficient determines whether the home country's FG increases the foreign output gap. The BOJ implemented a ZIRP in February 1999, but decided to end the policy in August 2000. Several economists have criticized the BOJ for once again facing deflationary risks as a result of the ZIRP's premature termination. However, our simulation results indicate that even adopting a home country's longer-length FG does not adequately boost inflation and the output gap from an open economy perspective. If the home central bank implements a reinforced FG strategy, the home output gap will steadily increase. However, if the CRRA coefficient is 2.0, a home country FG may reduce foreign social welfare. As a result, although the empirically plausible value of the CRRA coefficient is debatable in an open economy, the home central bank's ability to extend the FG is limited if the natural rate of interest declines only in the home country.

\subsection{International FG interaction under a global liquidity trap}
Consider the second scenario and the primary focus of this study: a simultaneous global liquidity trap. We examine the FG interaction between two countries in a situation where the home and foreign economies are both constrained by the ZLB as a result of a global decline in the natural rate of interest. Following the financial crisis that began in the US economy, central banks in advanced economies set nominal interest rates to zero. Furthermore, the recent COVID-19 shock triggered a severe global recession, forcing central banks to deal with both ZLB constraints and the aftermath. Except for the BOJ's policy rate, central banks in advanced economies began raising interest rates after the COVID-19 shock subsided. These point to episodes of a global liquidity trap. Thus, taking into account the cross-country interaction effects of forward guidance would aid our understanding of the practical aspects of the central bank suggested by the preceding evidence.

\begin{figure}[h]
\caption{The impulse response to a global liquidity trap shock: Country F with the ZLB constraint.}
\includegraphics[width=14cm,height=14cm]{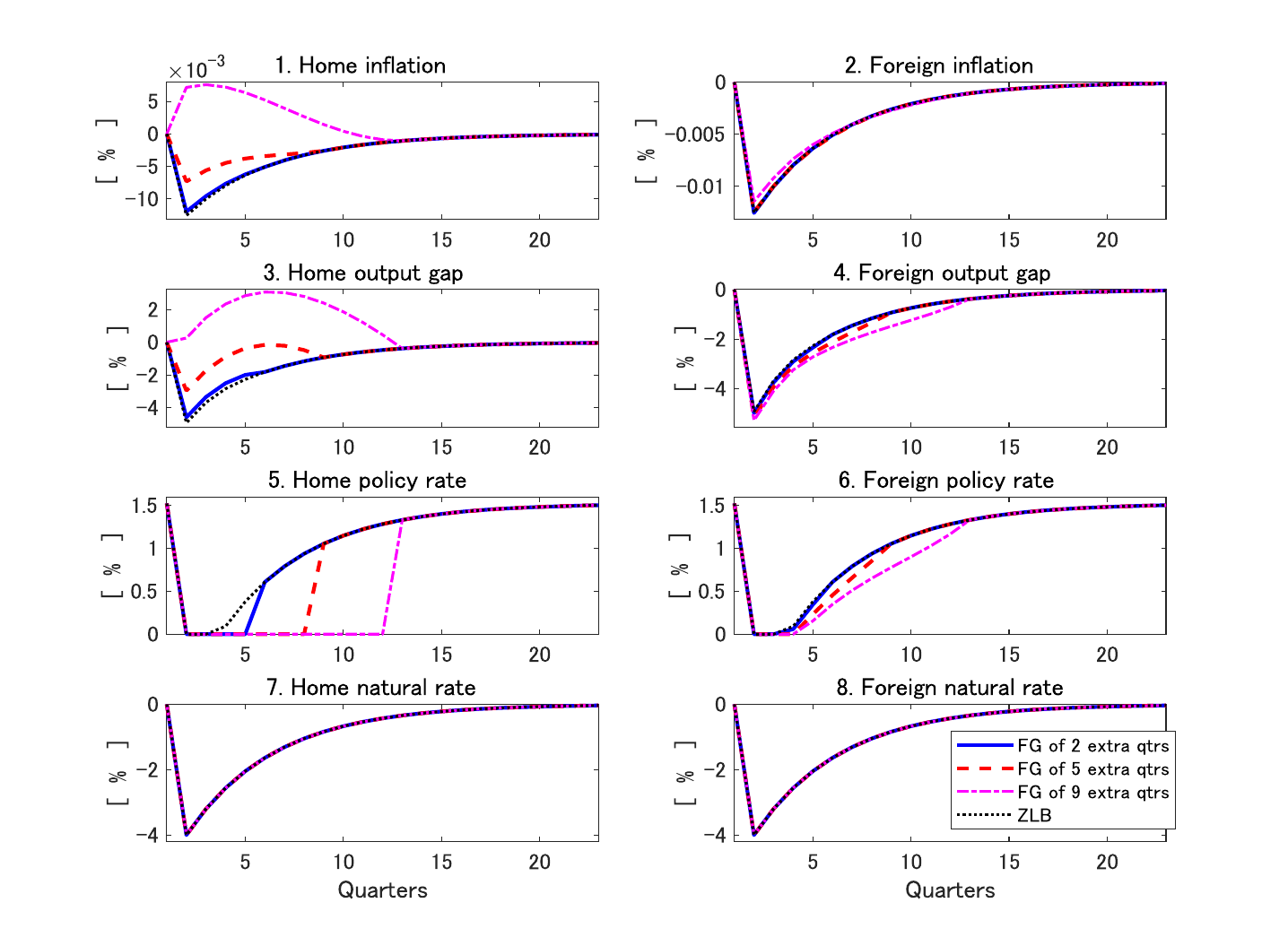}
\begin{flushleft}
\end{flushleft}
\end{figure}

Figure 5 presents the scenario in which only the home central bank adopts FG in the event of a global liquidity trap. In contrast to Figure 2, when the natural rate of interest falls simultaneously in both countries, the inflation rate and output gap fall in both the home and foreign economies. If the home and foreign central banks do not adopt the FG, the zero interest rate period ends on the same date in both countries. As long as the home central bank uses a short fixed-length FG, it is unable to stimulate inflation and the output gap in its own country.

However, if the home central bank adopts a five-period fixed-length FG, it can mitigate the decline in the inflation rate and output gap caused by negative demand shocks. In contrast, in a foreign country, even if the central bank extends the termination of its ZIRP, such a policy prescription is insufficient to raise inflation and the output gap. Surprisingly, introducing a nine-period fixed-length FG in the home country allows the home central bank to successfully boost inflation and the output gap. In contrast to the home country, the foreign central bank is still experiencing a severe recession due to a global liquidity trap. To summarize, in a global liquidity trap situation, the country where the central bank establishes a long-period fixed-length FG avoids a severe recession caused by a decrease in the natural interest rate.

Does a long-term fixed-length FG adopted by one country's central bank improve global welfare? Table 3(a) shows that when the home central bank does not adopt FG, the foreign country experiences the smallest welfare loss. In contrast, adopting a five-period fixed-length FG results in the smallest welfare loss for the home country. However, the home central bank's nine-period fixed-length FG policy worsens the home country's social welfare. This is because long-period FGs have an additional stimulus effect on inflation and the output gap, increasing their volatility.

\include{Table_3.tex}

Figure 6 shows the impulse response to the global liquidity trap when the home and foreign central banks use the same fixed-length FG. If central banks in both countries do not use FG after a global liquidity trap shock, as shown in Figure 5, a severe recession will result from a decline in the natural interest rate. Figure 6 shows that a two-period fixed-length FG adopted by home and foreign central banks would be insufficient to overcome a severe recession in each country. Unlike the two-period fixed-length FG, both central banks can reduce inflation and the output gap in each country by implementing the five-period fixed-length FG simultaneously. Surprisingly, when home and foreign central banks try to implement a nine-period fixed-length FG, they immediately succeed in stimulating the economy. According to Table 3(b), adopting a six-period fixed-length FG simultaneously by both home and foreign central banks reduces global welfare losses. Clearly, such home and foreign prescriptions reduce global welfare losses.

\begin{figure}[h]
\caption{The impulse response to a global liquidity trap shock: Both central banks adopt the same length of FG}
\includegraphics[width=14cm,height=14cm]{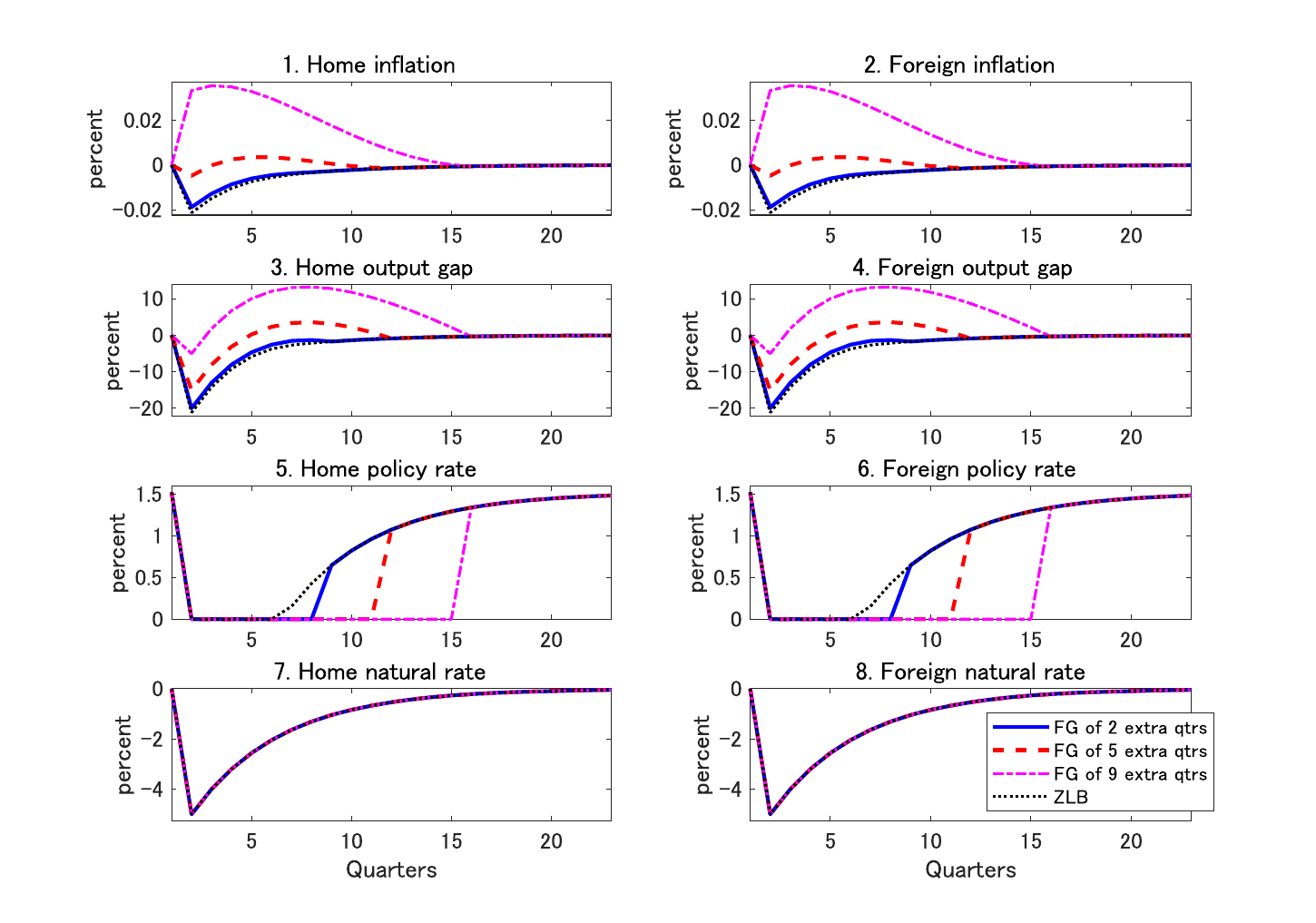}
\begin{flushleft}
\end{flushleft}
\end{figure}

Finally, suppose the foreign central bank always uses a five-period fixed-length FG and the home central bank has the freedom to choose the length of the FG. Figure 7 depicts this case. If only the foreign central bank implements a five-period fixed-length FG, both central banks will experience a severe recession. This result corresponds to the exact opposite case of Figure 5. Even if the home central bank considers a two-period fixed-length FG, it will still face a severe recession due to a global liquidity trap. Importantly, if the home central bank implements a nine-period fixed-length FG, it is successful in increasing inflation and the output gap in its own country. However, even after implementing a five-period FG, the foreign central bank is unable to overcome the recession. Thus, if a foreign central bank adopts fixed-length FG, it means that the home central bank can stimulate inflation and the output gap if its fixed-length FG policy is longer than the foreign central bank's.

\begin{figure}[h]
\caption{The impulse response to a global liquidity trap shock: Country F adopts fixed 5 extra quarters FG}
\includegraphics[width=14cm,height=14cm]{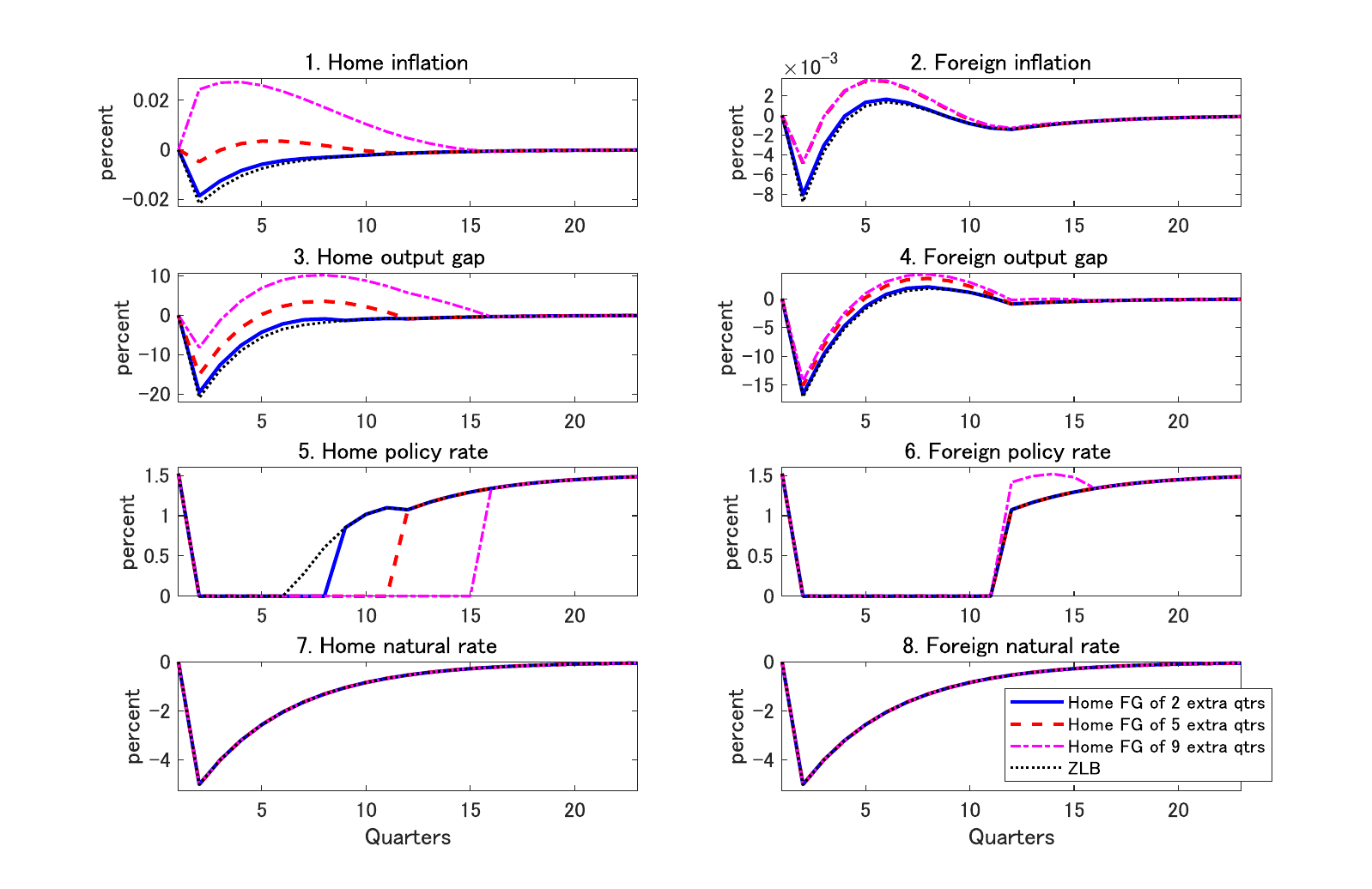}
\begin{flushleft}
\end{flushleft}
\end{figure}

Table 3(c) shows that a foreign central bank can achieve the smallest welfare loss by implementing a five-period fixed-length FG while the home central bank maintains its monetary policy conduct without FG. Evidently, this is because the foreign country can benefit from the expenditure-switching effect of an exchange rate depreciation. Implementing a nine-period fixed-length FG by the home central bank can reduce the foreign country's expenditure-switch effect, resulting in a boom. This finding suggests that, while a longer fixed-length FG in the home country raises inflation and the output gap, it may also result in an additional boom. Conversely, foreign central banks cannot stimulate inflation and the output gap effectively. The combination of a home country's nine-period fixed-length FG and a foreign country's five-period fixed-length FG produces unfavorable results, as shown in Table 3(c). Accordingly, we conclude that when domestic and foreign central banks use the same length of FG, the global welfare loss is minimized.

In light of the aforementioned, in scenarios where each country's primary goal is to maximize its own welfare, a shift in the foreign central bank's monetary policy stance following the COVID-19 crisis may conflict with the home country's FG policy. These conflicts may arise as a result of a lack of alignment regarding the mutual benefits inherent in each country's monetary policies, potentially creating an incentive to deviate from implicit international monetary policy coordination. As previously stated, while a detailed simulation of the costs of deviating from international monetary policy coordination is beyond the scope of this study, the following section presents a simulation that as closely as possible replicates the FRB and BOJ's different monetary policy stances following the COVID-19 crisis. This examination is motivated by the FRB's departure from the ZIRP, which was intended to mitigate the US economy's inflation surge following the COVID-19 pandemic, and how it affects the BOJ's monetary policy strategy.

\section{Discussion}
This section discusses the policy implications of FG policies' interaction effect on international monetary policy transmission based on our main findings. We also discuss the FG puzzle's caveats and limitations in this study. 

\subsection{Policy implication}
In a two-country NK model, this study explored how FGs interact internationally. The following are the main findings of our study. First, unlike when the home central bank does not use FG, introducing the longer fixed-length FG can boost inflation and the output gap in its own country. However, a stronger FG policy may result in an additional boom in the home country. In contrast, the foreign country is experiencing a severe recession as a result of a global liquidity trap shock. Second, even if the foreign central bank adopts a specific length of FG, the home central bank can increase inflation and the output gap in its own country if its FG is longer than the foreign central bank's. In either case, the home central bank's FG may have a beggar-thy-neighbor effect on the foreign country. As a result, the combination of home and foreign FG policies negatively affects global social welfare. Note that whether the foreign country can benefit from the interaction effect of home and foreign FGs hinges on the value of the CRRA coefficient.

These assertions are in stark contrast with the results obtained in \citet{fujiwara2013global} and \citet{haberis2020welfare}.
For conventional monetary policy, several studies have pointed out an international spillover effect of QE \citep{bhattarai2021effects, kolasa2020international}. \citet{kolasa2020international} theoretically show that a large country's QE has a positive effect on its own GDP but a negative GDP effect on emerging countries. They argue that this result is consistent with recent empirical research on the international spillover path of QE. \citet{ida2023effect} examined the effect of real money balances on monetary policy in a two-country NK model and demonstrated that when the home and foreign countries face the ZLB, a foreign QE shock may have a beggar-thy-neighbor effect on the home country. Our study, on the other hand, focuses on the international spillover effects of FG policies in a two-country NK model using a simple Taylor rule. As far as we know, no studies address the normative policy implications for FG policy in a two-country NK model.

What implications does this study have for international monetary policy in terms of the interaction of FG policies? Inflationary pressures in the United States have prompted the Federal Reserve to tighten monetary policy sharply since 2022. Although it appeared that the FRB took a long time to subdue inflation, the US economy has not entered a deep recession as a result of the sharp rise in interest rates. In contrast, the BOJ's termination of its ZIRP took much longer than the US, widening the US-Japan interest rate differential. As a result, the Japanese yen has depreciated significantly against the dollar since 2022. In fact, Figure 8 shows that the yen has depreciated significantly against the US dollar, coinciding with a sharp widening of the Japan-US interest rate differential. Rising resource prices also causes inflationary pressures in Japan. However, the BOJ has been unable to increase policy rates, leaving it with no tools to combat inflationary pressures. Although it is difficult to calculate the welfare losses caused by the FRB's earlier termination of its FG policy than the BOJ's, it is possible that the BOJ's policy implementations were less flexible during a liquidity trap. Furthermore, the sharp depreciation of the yen against the dollar would not have hit the Japanese economy if the BOJ had terminated its ZIRP at the same time as the Fed's ZIRP termination. In this sense, our results may contradict the preceding fact.

\begin{figure}[h]
\begin{center}
\caption{Interest rate differential and the nominal exchange rate: Japan and U.S.}
\includegraphics[width=12.0cm,height=8cm]{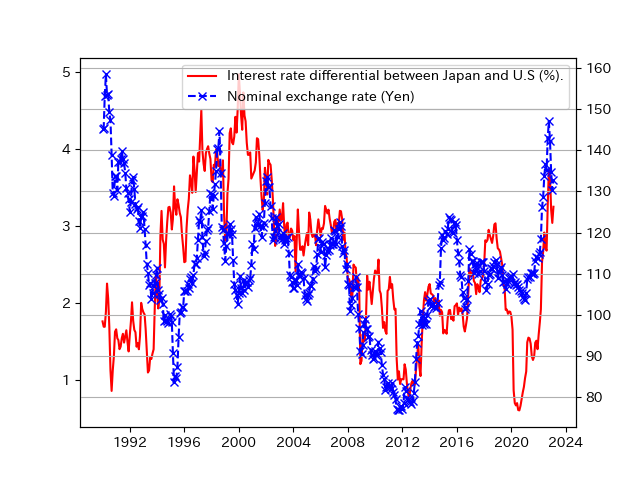}

Source: Federal Reserve Bank of St.Louis.
\end{center}
\end{figure}

In what follows, we perform a numerical exercise to capture the most recent situation between the Japanese and US economies in our model. First, we assume that foreign monetary policy does not follow the FG and remains in a ZLB situation. Under these conditions, the first period faces a global natural rate of interest shock, which is interpreted as a global crisis caused by a pandemic in 2020. The magnitude of this shock is the same as stated in Section 4. We also consider that the home country's monetary policy can follow one of five patterns, as shown in Table 3. Under these conditions, we present the case for a cost-push shock in a foreign country after five periods, which corresponds to inflationary pressures in the US economy in 2021.

Table 4 summarizes the simulation results, which show that including a cost-push shock worsens the welfare losses of both countries when compared to the case without the shock as a baseline. The adoption of a FG policy by one's own country exacerbates welfare losses in the other country. Thus, our calculation results imply that if we interpret the termination of the BOJ's negative interest rate policy in 2024, which was implemented following the termination of the FRB's ZIRP, as a modification of the BOJ's FG, the possibility of worsening welfare losses in home and foreign economies cannot be excluded.


\include{Table_4.tex}

Table 4(a) deals with cases where the CRRA coefficient is 2.0, whereas Tables 4(b) and 4(c) deal with cases where it is 1 or 0.5, respectively. As shown in Table 4(a), the home central bank's use of the five-period length FG reduces the home country's welfare losses. In contrast, the home central bank's four-period fixed-length FG reduces foreign welfare losses. Thus, the minimization of welfare losses in each country does not always correspond to the intention of each country's monetary policy. Moreover, in the case of $\sigma>1$, it is clear that home countries such as Japan have an incentive to choose longer fixed-length FG periods than other countries such as the United States would expect. The simulation results appear to reflect the BOJ and FRB's actual monetary policy stance during this period, as the results of Table 4(a) suggest that the continuation of the BOJ's ZIRP may have reduced US losses through the international FG transmission. In contrast, Japan's gains may last longer than expected, indicating that the interest rate gap between the two countries is widening as a result. 

Conversely, as Table 4(c) shows, in the case of $\sigma<1$, when the nine-period fixed-length FG is implemented, the foreign country's losses are reduced, while the home country's losses are increased during those periods. Unlike the previous case, this case shows that home countries, like Japan, may have a reason to choose shorter FG periods than other countries, like the United States, would expect. However, as shown in Table 4(b), in the case of $\sigma=1.0$, the minimization of losses for the two countries with the five-period fixed length FG appears to coincide, indicating that there is no conflict of interest between them. While the empirical plausibility of the CRRA coefficients is debatable, these results highlight the importance of taking international FG policy interactions into account. Adopting foreign FGs will become an important policy agenda in terms of international coordination during a severe global economic crisis, but we will not go into detail about this in this paper.

\subsection{Caveat and limitation}
We discuss the policy implications of the FG policy's impact on the international monetary transmission mechanism in a two-country NK model. While this paper provides several policy implications for the international interaction of FG policies, the FG effect obtained in this study may be stronger or weaker than that obtained in a closed economy. Several studies have argued that the effect of FG in the standard NK model outweighs the actual policy effects. For instance, \citet{carlstrom2015inflation} showed that when the duration of an FG policy exceeds a threshold value, the effect increases both inflation and the output gap. \citet{del2012forward} labelled this inconsistency as the FG puzzle. In particular, \citet{mckay2016power} pointed out that the extension of the central bank's commitment to a zero interest rate makes this FG puzzle more severe. While the NK model theoretically supports the effectiveness of the FG policy as a tool for non-conventional monetary policy, its power appears to be problematic. Several recent studies have focused on how to solve the FG puzzle by extending the standard NK model.\footnote{\citet{bodenstein2012imperfect} and \citet{haberis2019uncertain} showed that the power of an FG policy is weakened if the central bank's commitment to the FG policy is not entirely credible. \citet{boneva2018threshold} showed that a threshold-based FG policy outperforms a calendar-based one. \citet{campbell2019limits} showed the limitation of an FG policy and argued that an imperfect communication strategy of the central bank causes macroeconomic fluctuations.} To do this, \citet{del2012forward} incorporated the perpetual-youth model into the standard NK model. Furthermore, \citet{gabaix2020behavioral} showed that the FG puzzle is solved when bounded rationality is considered in the standard NK model.\footnote{\citet{nakata2019attenuating} examined the effect of an FG policy in a model where the forward-looking structural equations, such as the new Keynesian Phillips curve and the dynamic IS curve, are discounting. \citet{nakata2019attenuating} showed that in the case of an attenuated FG effect, the central bank can implement optimal monetary policy by temporarily extending the ZIRP.} \citet{mckay2016power} attenuated the effect of FG policy by incorporating incomplete markets into the standard NK model.

Furthermore, we would like to mention the limitations of our simulation results. We used the OccBin toolkit to create simulation results for a global liquidity trap. When a stochastic shock is present, the two-country model may complicate the numerical algorithm for solving the ZLB problem due to the presence of state variables. As a result, including a stochastic shock in the model may change the characteristics of the FG policy. For example, we can use a stochastic shock to reconsider the nonlinear effects of the ZLB in our model.

While these caveats and limitations are still important topics, extending the current model to include the aforementioned points is beyond the scope of this study. Of course, we acknowledge the limitations of comparing our results with those of previous studies, but we emphasize the importance of considering the role of international spillovers of FG policies.

\section{Conclusions}
This paper studied the effect of FG on international monetary transmission in a two-country NK model. While the effectiveness of FG has been discussed in the closed-economy NK model, it is still unclear how it affects the international transmission mechanism of monetary policy. In what ways do the effects of FG in one country affect other countries? Previous literature does not fully explain this possibility. Our study filled a gap in the canonical two-country NK model. Furthermore, following the turmoil of the financial crisis that began in the United States in 2008, it appears that central banks in advanced countries adopted the ZIRP as one of their unconventional monetary policies. Recently, the large shocks associated with the COVID-19 pandemic forced central banks to reintroduce the ZIRP to help the economy recover from a severe recession. These facts encouraged this study.

The three main findings of this study are as follows. First, the magnitude of the CRRA parameter is important in determining the beggar-thy-neighbor and prosper-thy-neighbor effects in foreign economies when the home country only faces the ZLB. In other words, when the CRRA coefficient is less than one, home FG has a negative economic effect on the foreign country with a symmetrical economic structure. In contrast, if the CRRA coefficient is greater than unity, home FG has the opposite economic effect on its counterpart country. Second, both countries may benefit from adopting only the home country's FG policies when the home central bank only faces the ZLB. Furthermore, extending the duration of FG over its counterpart will result in a duration of FG that minimizes the home country's welfare loss. Third, we address the potential benefit of the FG interaction effect between two countries. Thus, we argue the possibility that by adopting the same duration of FG quarters, home and foreign central banks can benefit from monetary policy coordination.
 
Finally, we would like to mention some future extensions to this study. As previously stated, we are unable to explicitly determine the optimal length of the FG policy under international policy coordination because our analysis is based on a numerical exercise. In the future, we would like to investigate how the optimal FG policy is analytically determined in a two-country framework. Furthermore, we use the case in which home and foreign central banks jointly choose the same size of FG policy to show the potential benefit of policy coordination. This result is based on the assumption that domestic and foreign central banks that follow a simple policy rule with FG choose the same size of FG policies. However, strictly speaking, we need to check whether our results correspond to the optimal cooperative policy, which implies that home and foreign central banks jointly minimize the worldwide loss function.

\bibliographystyle{elsestyle}
\bibliography{ref}

\include{appendix.tex}

\end{document}

%% file: Table_1.tex
\begin{table}[h]
\caption{Calibrated parameters}
\begin{center}
\begin{tabular}{clll}
\hline\hline
Parameter & Description& Value & \\ \hline
$\alpha$ & Calvo pricing for home country &    0.9 \\
$\alpha^*$  &  Calvo pricing for foreign country&   0.9  \\
$\beta$    & Discount factor &  0.985 \\
$\sigma$   & Relative risk aversion coefficient & 2.0  \\
$\gamma$    & Degree of openness &      0.5  \\
$\eta$        & Inverse of Frish labor elasticity & 1.5 \\
$\theta$     & Elasticity of substitution for individual goods &  10.0  \\
$\psi_r$    & Interest rate smoothing in the Taylor rule for Country H &  0  \\
$\psi_{\pi}$  & Inflation stabilization in the Taylor rule for Country H &  1.25  \\
$\psi_x$   & Output gap stabilization in the Taylor rule for Country H & 0.5 \\
$\psi^*_r $   & Interest rate smoothing in the Taylor rule for Country F  & 0 \\
$\psi^*_{\pi}$  & Inflation stabilization in the Taylor rule for Country F & 1.25  \\
$\psi^*_x $   & Output gap stabilization in the Taylor rule for Country F & 0.5  \\
$\rho_r$    & AR(1) coefficient for natural interest rate shock in Country H &  0.8  \\
$\rho^*_r $    & AR(1) coefficient for natural interest rate shock in Country F  & 0.8 \\
$e_t^{NR}$  & Size of shock to natural rate of interest in Country H & -0.05 \\
$e_t^{GL}$  & Size of shock to global liquidity trap in Country H & -0.04 \\
$(e_t^{GL})^*$  & Size of shock to global liquidity trap in Country F & -0.04\\ \hline
\end{tabular}%
\end{center}
\end{table}

%% file: Table_2.tex
\begin{table}[h]
\caption{Welfare losses in a natural rate shock of only Country H}
\begin{center}
(a) High degree of CRRA ( $ \sigma = 2.0 $ ) \\
\begin{tabular}{ccccc}
\hline\hline
H policy & F policy    &    World loss    &   H loss   & F  loss  \\ \hline
ZLB      &  no ZLB   &   23.94   &   23.50   &    0.44   \\
FG 2 extra qtrs  &  no ZLB  & 22.03 &   21.64 & 0.39 \\
FG 4 extra qtr  &  no ZLB & 17.30 &  16.97 &  0.33 \\
FG 5 extra qtr &  no ZLB &  $\bf{14.77}^{\star}$ &  $\bf{14.46}^{\star}$ & $\bf{0.31}^{\star}$   \\
FG 9 extra qtr  &  no ZLB & 19.94 &  19.42 & 0.52\\
FG 10 extra qtr &  no ZLB &  29.72 & 29.04 & 0.68  
\\ \hline
\end{tabular}%
\end{center}

\begin{center}
(b) Medium degree of CRRA ( $ \sigma = 1.0 $ ) \\
\begin{tabular}{ccccc}
\hline\hline
H policy & F policy    &    World loss    &   H loss   & F  loss  \\ \hline
ZLB      &  no ZLB   &   26.95  &  26.95  &     0.00 \\
FG 2 extra qtr  &  no ZLB  & 25.30 &      25.30  &      0.00 \\
FG 4 extra qtr  &  no ZLB & 19.26 &    19.26  &      0.00 \\
FG 5 extra qtr  &  no ZLB &  $\bf{15.84}^{\star}$ &   $\bf{15.84 }^{\star}$  &    0.00   \\
FG 9Q Extra  &  no ZLB & 24.54 &   24.54   &    0.00
\\ \hline
\end{tabular}%
\end{center}

\begin{center}
(c) Low degree of CRRA ( $ \sigma = 0.5 $ ) \\
\begin{tabular}{ccccc}
\hline\hline
H policy & F policy    &    World loss    &   H loss   & F  loss  \\ \hline
ZLB      &  no ZLB   &  33.27  &  32.67  &   0.60   \\
FG 2 extra qtr  &  no ZLB  & 28.28 &  27.72 &  0.57 \\
FG 4 extra qtr  &  no ZLB & 19.01  &  18.44 &  $\bf{0.56}^{\star}$ \\
FG 5 extra qtr &  no ZLB &  $\bf{15.65}^{\star}$&  $\bf{15.06 }^{\star}$ &  0.59  \\
FG 9 extra qtr  &  no ZLB & 55.31  &  54.18 &  1.13
\\ \hline
\end{tabular}%
\end{center}
\begin{flushleft}
\small{Note: The H policy column represents the home country's monetary policy, whereas the F policy column represents the foreign country's monetary policy. No ZLB indicates monetary policy without the normal linear model's ZLB constraint, whereas ZLB indicates the zero lower bound constraint.}
\end{flushleft}
\end{table}

%% file: Table_3.tex
\begin{table}[h]
\caption{Welfare losses under a global liquidity trap: Case of High degree of CRRA ( $ \sigma = 2.0 $ )}
\begin{center}
(a) Country F with ZLB constraint \

\begin{tabular}{ccccc}
\hline\hline 
H policy & F policy    &    World loss    &   H loss   & F  loss  \\ \hline
ZLB      &   ZLB   & 27.20   &   13.60  &    $\bf{13.60}^{\star}$  \\
FG 2 extra qtr  &  ZLB  & 26.94 &   13.19  &    13.75  \\
FG 4 extra qtr  &  ZLB & 25.38 &   11.27  &    14.11 \\
FG 5 extra qtr &  ZLB &  $\bf{24.33}^{\star}$ &   $\bf{10.06}^{\star}$  &    14.27   \\
FG 10 extra qtr &  ZLB &  35.59 &   21.80 &     13.79  
\\ \hline
\end{tabular}%
\end{center}

\begin{center}
(b) Both central banks adopt the same length of FG
\begin{tabular}{ccccc}
\hline\hline 
H policy & F policy    &    World loss    &   H loss   & F  loss  \\ \hline
ZLB      &  ZLB   &    27.20  &    13.60   &   13.60   \\
FG 2 extra qtr  &  FG 2 extra qtr  &  26.73   &   13.37  &    13.37   \\
FG 4 extra qtr  &  FG 4 extra qtr & 23.82   &   11.91    &  11.91  \\
FG 6 extra qtr &  FG 6 extra qtr &  $\bf{19.86}^{\star}$   &    $\bf{9.93}^{\star}$   &    $\bf{9.93}^{\star}$ \\
FG 10 extra qtr &  FG 10 extra qtr &  33.00  &    16.50  &    16.50
\\ \hline
\end{tabular}%
\end{center}

\begin{center}
(c) Country F adopt fixed five extra quarters FG 

\begin{tabular}{ccccc}
\hline\hline 
H policy & F policy    &    World loss    &   H loss   & F  loss  \\ \hline
ZLB      &  FG 5 extra qtr   &  24.33   &   14.27  &    $\bf{10.06}^{\star}$  \\
FG 5 extra qtr &  FG 5 extra qtr &  21.72  &   10.86  &    10.86   \\
FG 6 extra qtr  &  FG 5 extra qtr & $\bf{20.83}^{\star}$  &   $\bf{9.62}^{\star}$  &    11.20 \\
FG 10 extra qtr &  FG 5 extra qtr &  32.89 &   20.79  &    12.10 
\\ \hline
\end{tabular}%
\end{center}
\begin{flushleft}
\small{Note: The H policy column represents the home country's monetary policy, while the F policy column represents the foreign country's monetary policy. No ZLB indicates monetary policy without the normal linear model's ZLB constraint, whereas ZLB indicates the zero lower bound constraint. The CRRA coefficient has a fixed value of 2.}
\end{flushleft}
\end{table}

%% file: Table_4.tex
\begin{table}[h]
\caption{Welfare losses when Country F's cost-push shock occurs under a global liquidity trap}
\begin{center}
(a) High degree of CRRA ( $ \sigma = 2.0 $ ) \\
\begin{tabular}{ccccc}
\hline\hline 
H policy & F policy    &    World loss    &   H loss   & F  loss  \\ \hline
ZLB      &   ZLB   & 93.11   &   21.02  &    72.09  \\
FG 2 extra qtr  &  ZLB  & 92.18  &  20.05   &   72.14  \\
FG 4 extra qtr  &  ZLB &  59.49  &  11.29 &   $\bf{48.20}^{\star}$  \\
FG 5 extra qtr &  ZLB &  $\bf{55.18}^{\star}$  &  $\bf{6.76}^{\star}$  &    48.42    \\
FG 9 extra qtr &  ZLB &  66.73 &  14.70 & 52.03  
\\ \hline
\end{tabular}%
\end{center}

\begin{center}
(b) Medium degree of CRRA ( $ \sigma = 1.0 $ ) \\
\begin{tabular}{ccccc}
\hline\hline 
H policy & F policy    &    World loss    &   H loss   & F  loss  \\ \hline
ZLB      &   ZLB   & 100.88  &  26.95 & 73.93   \\
FG 2 extra qtr  &  ZLB  & 99.22 & 25.30 & 73.93 \\
FG 4 extra qtr  &  ZLB &  62.30 & 12.14 & $\bf{50.16}^{\star}$ \\
FG 5 extra qtr &  ZLB &  $\bf{60.00}^{\star}$  & $\bf{9.85}^{\star}$ & $\bf{50.16}^{\star}$   \\
FG 9 extra qtr &  ZLB &  84.75 & 34.59 &  $\bf{50.16}^{\star}$ 
\\ \hline
\end{tabular}%
\end{center}

\begin{center}
(c) Low degree of CRRA ( $ \sigma = 0.5 $ ) \\
\begin{tabular}{ccccc}
\hline\hline 
H policy & F policy    &    World loss    &   H loss   & F  loss  \\ \hline
ZLB      &   ZLB   & 119.11   &   44.76  &    74.35   \\
FG 2 extra qtr  &  ZLB  & 114.07  &  39.88 &  74.19  \\
FG 4 extra qtr  &  ZLB &  64.48  &  14.65 &  49.83 \\
FG 5 extra qtr &  ZLB &  $\bf{61.21}^{\star}$ &  $\bf{12.85}^{\star}$ &  48.37     \\
FG 9 extra qtr &  ZLB &  92.64  & 47.78 & $\bf{44.86}^{\star}$ 
\\ \hline
\end{tabular}%
\end{center}
\begin{flushleft}
\small{Note: The H policy column represents the home country's monetary policy, while the F policy column represents the foreign country's monetary policy. No ZLB indicates monetary policy without the normal linear model's ZLB constraint, whereas ZLB indicates the zero lower bound constraint.}
\end{flushleft}
\end{table}

%% file: appendix.tex
\begin{appendix}
\section{Technical Appendix}
\subsection{Model}
\renewcommand{\theequation}{A.\arabic{equation} }
\setcounter{equation}{0}
In this note, we derive a two-country new Keynesian model between the two countries. This model is based on the framework developed by \citet{clarida2002simple}. We consider an economy with two symmetric large countries, a home country and a foreign country. The sizes of the home and foreign economies are $1-\gamma$ and $\gamma$, respectively.

There are two production sectors in each country. The final goods sector, which is characterized by perfect competition. The intermediate goods sector faces monopolistic competition and \citet{calvo1983staggered} type nominal price rigidities. We acknowledge that the degree of price stickiness varies across countries. The number of final goods producers is equal to the number of households in each country. We also assume that there is a complete market in both countries and that only final goods are traded. The case of producer currency prices is assumed, which implies complete pass-through of the exchange rate.

Finally, unless otherwise noted, similar equations hold for foreign countries. Also, note that foreign variables are denoted with an asterisk.

\subsubsection{Households}
\textbf{Preferences}
Preferences for consumption in the home country are given by
\begin{align}
C_t \equiv C_{H,t}^{1-\gamma}C_{F,t}^{\gamma}, \label{eq:1}
\end{align}
where $C_{H,t}$ is the consumption of domestic goods and $C_{F,t}$ is the consumption of foreign goods. The price index in the home country is given by:
\begin{align}
P_t=k^{-1}P_{H,t}^{1-\gamma}P_{F,t}^{\gamma} =k^{-1}P_{H,t}S_t^{\gamma}, \label{eq:2}
\end{align}
where $k \equiv (1-\gamma)^{(1-\gamma)}\gamma^{\gamma}$, $P_{H,t}$ is the price of domestic goods and $P_{F,t}$ is the price of foreign goods. Also, $S_t$ represents the terms of trade, which is given by
\begin{align}
S_t \equiv \frac{P_{F,t}}{P_{H,t}}. \label{eq:3}
\end{align}

\textbf{Household's optimization problem}
The intertemporal utility of an infinitely lived representative household is
\begin{align}
E_0 \sum_{t=0}^{\infty}\beta^t U_t=E_0 \sum_{t=0}^{\infty}\beta^t \bigg \{u\bigg(C_t, \frac{M_t}{P_t} \bigg)- V(N_t) \bigg \}, \nonumber
\end{align}
where $C_t$ is consumption and $N_t$ is the household's labor supply. We assume that the utility function, $u(\cdot)$, is strictly concave and continuously differentiable, and the disutility of labor supply, $V(\cdot)$, is strictly convex and continuously differentiable. We assume that the utility function of households is separable into consumption and real money balances.

The representative household maximizes the above utility function subject to the following budget constraint:
\begin{align}
P_tC_t+M_t+E_t[Q_{t,t+1}B_{t+1}]=B_t+M_{t-1}+W_tN_t+\Gamma_t-T_t, \nonumber
\end{align}
where $B_t$ is nominal bonds held for one period, $M_t$ denotes nominal money supply, and $W_t$ and $\Gamma_t$ are the nominal wage and dividend, respectively, earned from domestic firm. Also, $T_t$ denotes the lump-sum tax.

We assume that a complete market is present in both countries, and introduce the following stochastic discount factor:
\begin{align}
E_t(Q_{t,t+1})=\frac{1}{1+r_t}, \label{eq:4}
\end{align}
where $Q_{t,t+1}$ denotes a stochastic discount factor and $r_t$ is the risk free short-term nominal interest rate. 

We assume that the purchasing power parity condition holds for this economy: 
\begin{align}
P_t = \mathcal{E}_t P_t^{*}, \label{eq:5}
\end{align}
where $\mathcal{E}_t$ is the nominal exchange rate and $P_t^{*}$ is the price level in the foreign country. 

The first order conditions of this household's optimization problem are as follows:
\begin{align}
Q_{t,t+1}=\beta\frac{u_c(C_{t+1},Z_{t+1})}{u_c(C_t,Z_t)}\frac{P_t}{P_{t+1}}, \label{eq:6}
\end{align}
\begin{align}
\frac{u_m(C_t, Z_t)}{u_c(C_t, Z_t)}=\frac{r_t}{1+r_t}, \label{eq:8}
\end{align}
\begin{align}
-\frac{V_n(N_t)}{u_c(C_t,Z_t)}=\frac{W_t}{P_t}, \label{eq:9}
\end{align}
where $Z_t = M_t/P_t$ denotes real money balances. 

Taking the expectation for Eq.(\ref{eq:6}),
\begin{align}
E_t[Q_{t,t+1}]=\frac{1}{1+r_t}=\beta E_t\bigg[\frac{u_c(C_{t+1},Z_{t+1})}{u_c(C_t,Z_t)}\frac{P_t}{P_{t+1}} \bigg]. \label{eq:7}
\end{align}
In the subsequent discussion, we assume a separable utility function between consumption and real money balances.

\subsubsection{International risk-sharing}
Next, we consider a risk-sharing condition between countries. The Euler equation for foreign consumption denominated in home currency is
\begin{align}
\frac{1}{1+r_t^*}=\beta E_t\left[\frac{u_c(C^*_{t+1},Z^*_{t+1})}{u_c(C^*_t,Z^*_t)}\frac{P^*_t \mathcal{E}_t}{P^*_{t+1} \mathcal{E}_{t+1}}\right]. \label{eq:b10}
\end{align}
As in \citet{clarida2002simple}, we assume that the first order conditions are symmetric across countries and the power parity condition holds. Under the separable utility function between consumption and real money balances, as shown in \citet{clarida2002simple}, we obtain the following result:
\begin{align}
C_t=C^*_t,  \label{eq:b11}
\end{align}
for all $t$.

\subsubsection{Firms}

\textbf{Final goods firm}

The final goods sector is perfectly competitive and producers use inputs that are produced in the intermediate goods sector. In particular, final goods are produced according to the following CES aggregate:
\begin{align}
Y_t=\left[\int_0^1 Y_t(i)^{\frac{\theta-1}{\theta}} di \right]^{\frac{\theta}{\theta-1}}, \label{eq:12}
\end{align}
where $Y_t$ is aggregate output, $Y_t(i)$ is demand for intermediate goods produced by firm $i$, and $\theta$ is the elasticity of substitution. Note that both variables are normalized by $1-\gamma$. 

Under the CES aggregate, the demand function is given by
\begin{align}
Y_t(i)=\left(\frac{P_{H,t}(i)}{P_{H,t}}\right)^{-\theta}Y_t,  \label{eq:13}
\end{align}
and the domestic price level is defined as:
\begin{align}
P_{H,t}=\left[\int_0^1 P_{H,t}(i)^{1-\theta}di \right]^{\frac{1}{1-\theta}}, \label{eq:14}
\end{align}
where $P_{H,t}(i)$ is the prices for intermediate goods produced by the firm $i$. Note that these variables are also normalized by $1-\gamma$.

\textbf{The intermediate goods sector}
The intermediate goods sector is characterized by monopolistic competition, and each firm produces a differentiated intermediate good. Firm $i$'s production function is given by
\begin{align}
Y_t(i)=A_tN_t(i), \label{eq:15}
\end{align}
where $A_t$ denotes an aggregate productivity disturbance.  

As in \citet{clarida2002simple}, the intermediate firm's real marginal cost is given as follows:
\begin{align}
\varphi_t=(1-\tau)\frac{W_t}{P_{H,t}}\frac{1}{A_t}. \label{eq:16}
\end{align}
Using the household's first order conditions, we can rewrite Eq. (\ref{eq:16}) as follows:
\begin{align}
\varphi_t=\frac{1-\tau}{k A_t}\frac{V_n(N_t)}{u_c(C_t,m_t)}S_t^{\gamma}. \label{eq:17}
\end{align}
Eq. (\ref{eq:17}) reveals that the home real marginal cost depends on the terms of trade in an open economy compared to the closed economy model.

Following \citet{calvo1983staggered}, we assume that price rigidity is present in the intermediate goods sector. The following explanation focuses on the home country. A fraction $1-\alpha$ of all firms adjusts their price while the remaining fraction of firms $\alpha$ do not. 

We now consider the intermediate firms that can adjust their price. When revising their prices, these firms take into account uncertainty concerning when they will be able to adjust prices next. In this case, the intermediate firm's optimization problem for the home country is given by
\begin{align}
E_t\sum_{t=0}^{\infty}(\alpha\beta)^j Q_{t,t+j}Y_{t+j}(i)(P_{H,t}^{opt}-P_{H,t+j}\varphi_{t+j}). \label{eq:19}
\end{align}
where $P_{H,t}^{opt}$ is the firm's optimal price. The first order condition of this maximization problem is as follows:
\begin{align}
E_t\sum_{t=0}^{\infty}(\alpha\beta)^j Q_{t,t+j}Y_{t+j}(i)(P_{H,t}^{opt}-(1+\mu) P_{H,t+j}\varphi_{t+j})=0. \label{eq:20}
\end{align}
where the variable $\mu=1/(\theta-1)$ is the price mark-up. In particular, when $\alpha=0$, this equation takes the following form:
\begin{align}
\frac{P_{H,t}^o(i)}{P_{H,t}} = (1+\mu)\varphi_t. \label{eq:21}
\end{align}
Finally, the price level in the intermediate goods sector is defined as:
\begin{align}
P_{H,t}=\left[\alpha (P_{H,t-1})^{1-\theta} +(1-\alpha) (P_{H,t}^{opt})^{1-\theta}\right]^{\frac{1}{1-\theta}}. \label{eq:22}
\end{align}

\subsubsection{Equilibrium}
We now describe the equilibrium conditions in an open economy. The equilibrium conditions for the goods market are given as follows:
\begin{align}
(1-\gamma )Y_t=(1-\gamma )C_{H,t}+\gamma C_{H,t}^{*}, \label{eq:23}
\\
\gamma Y_t^{*} = (1-\gamma )C_{F,t}+\gamma C_{F,t}^{*}. \label{eq:24}
\end{align}
Since we assume that the elasticity of substitution between home and foreign goods is one, purchasing power parity holds. In this case the real exchange rate is one:
\begin{align}
\frac{\mathcal{E}_t P^*_t}{P_t}=1. \nonumber
\end{align} 
Furthermore, under the assumptions that the consumption index follows a Cobb-Douglas specification and that the purchasing power parity condition holds, current accounts in both countries always equalize because the ratio of home income to foreign income is constant. As this implies that the trade balance is zero, the following conditions hold:
\begin{align}
P_{H,t}Y_t=P_tC_t, \label{eq:25}
\\
P^*_{F,t} Y^*_t=P^*_t C^*_t. \label{eq:26}
\end{align}
In turn, substituting Eq. (\ref{eq:25}) into Eq. (\ref{eq:2}), we obtain the following equation:
\begin{align}
Y_t=k^{-1}C_tS_t^{\gamma}. \label{eq:27}
\end{align}
At this point, the home terms of trade are represented by the ratio of home output to foreign output: 
\begin{align}
S_t=\frac{Y_t}{Y^*_t}. \label{eq:28}
\end{align}
Eq. (\ref{eq:28}) indicates that holding domestic output constant, an increase in foreign output leads to an appreciation of the home terms of trade.

On the other hand, due to complete risk-sharing in both countries, we also obtain the following equation:
\begin{align}
C_t=k(Y_t)^{1-\gamma}(Y^*_t)^{\gamma}. \label{eq:29}
\end{align}
According to Eq. (\ref{eq:29}), holding home output constant, a rise in foreign output induces an increase in home consumption. Home consumption increases less than a rise in home output because complete risk-sharing leads to consumption smoothing of households. Using the assumption of separable utility between consumption and real balances and substituting Eq. (\ref{eq:29}), we can rewrite Eq. (\ref{eq:17}) as follows:
\begin{align}
\varphi_t=\frac{1-\tau}{A_t}\frac{v_n(Y_t/A_t)}{u_c((Y_t)^{1-\gamma}(Y^*_t)^{\gamma})}\bigg(\frac{Y^*_t}{Y_t}\bigg)^{\gamma}. \label{eq:30}
\end{align}
It follows from Eq. (\ref{eq:30}) that the home real marginal cost depends not only on domestic output, but also on foreign output. For instance, from Eq. (\ref{eq:28}), the terms of trade improve when foreign output increases. The improvement in the terms of trade leads to a decline in the home real marginal cost. Consequently, the decline in home marginal cost induces a decrease in home inflation. This mechanism is referred to as the terms of trade externality. On the other hand, an increase in foreign output pushes the home real marginal cost up due to consumption risk-sharing between countries. As pointed out in \citet{clarida2002simple}, whether which of two effects dominates movements in the home real marginal cost depends on the value of the relative risk aversion coefficient for consumption.

\textbf{Equilibrium under flexible prices}
First of all, $\bar{H}$ represents the value of steady state, $H_t^n$ is the value of efficient level. Also, we define $h_t= \log(H_t/\bar{H})$ as the deviation of $H_t$ from steady state. The log-linearization of the efficient level of domestic output is given by
\begin{align}
&[\sigma+\eta-\gamma(\sigma-1)] y_t^n +\gamma(\sigma-1)y^{n*}_t=(1+\eta)a_t, \label{eq:31}
\\
&[\sigma+\eta-(1-\gamma)(\sigma-1)] y^{n*}_t +(1-\gamma)(\sigma-1)y_t^n=(1+\eta)a^*_t, \label{eq:32}
\end{align}
where $\sigma \equiv -u_{cc}\bar{C}/u_c$ and $\eta \equiv -V_{yy}\bar{y}/V_y$. $y_t^n$ denotes the efficient level of domestic output and $y^{n*}_t$ represents the efficient level of foreign output.

\textbf{Equilibrium under sticky prices: Log-linearilzation}

The structural equations in a two-country model are summarized as follows. In this derivation, we define the output gap for both countries as $x_t=y_t-y_t^n$ and $x^*_t=y^*_t-y^{n*}_t$, respectively.

\begin{itemize}
\item New Keynesian Phillips curve
\end{itemize}
\begin{align}
&\pi_t = \beta E_t \pi_{t+1} + \kappa_1 x_t + \kappa_2 x^*_t + u_t, \label{eq.hnkpc}
\\
&\pi^*_t = \beta E_t \pi^*_{t+1} + \kappa^*_1 x^*_t + \kappa^*_2 x_t + u^*_t. \label{eq.fnkpc}
\end{align}
\begin{itemize}
\item Dynamic Investment-Saving (IS) curve
\end{itemize}
\begin{align}
&x_t = E_t x_{t+1} +\vartheta[E_t\Delta x^*_{t+1}] -\sigma_0^{-1}(r_t - E_t \pi_{t+1}  - r_t^n), \label{eq.his}
\\
&x^*_t = E_t x^*_{t+1} +\vartheta^*[E_t\Delta x_{t+1}] -(\sigma^*_0)^{-1}(r^*_t - E_t \pi^*_{t+1}  - (r_t^n)^*). \label{eq.fis}
\end{align}
\begin{itemize}
\item Money demand function\footnote{The money demand function is redundant in this paper. As mentioned earlier, this is because the household's utility function is assumed to be separable between consumption and real money balances.}
\end{itemize}
\begin{align}
&m_t = \eta_y x_t + \eta_y^* x^*_t -\eta_r r_t, \label{eq.hmond}
\\
&m_t = \eta^*_y x^*_t + \eta_y x_t -\eta^*_r r^*_t. \label{eq.fmond}
\end{align}
where
\begin{align}
&\kappa_1 = \frac{(1-\alpha)(1-\alpha\beta)}{\alpha}(\sigma+\eta-\gamma(\sigma-1)), \nonumber \\
&\kappa_2 = \frac{(1-\alpha)(1-\alpha\beta)}{\alpha}\gamma(\sigma-1), \nonumber \\
&\kappa^*_1 = \frac{(1-\alpha^*)(1-\alpha^*\beta)}{\alpha^*}(\sigma+\eta-(1-\gamma)(\sigma-1)), \nonumber \\
&\kappa^*_2 = \frac{(1-\alpha^*)(1-\alpha^*\beta)}{\alpha^*}(1-\gamma)(\sigma-1), \nonumber \\
& \vartheta = \frac{\gamma(\sigma-1)}{\sigma-\gamma(\sigma-1)}, \vartheta^* = \frac{(1-\gamma)(\sigma-1)}{\sigma-(1-\gamma)(\sigma-1)}, \nonumber \\
& \sigma_0 = \sigma -\gamma(\sigma-1), \sigma^*_0 = \sigma -(1-\gamma)(\sigma-1). \nonumber
\end{align}

Finally, from the definition of the terms of trade, the nominal exchange rate evolves as follows:
\begin{align}
\epsilon_t = \epsilon_{t-1}+s_t-s_{t-1}+\pi_t-\pi_t^*, \label{eq:38}
\end{align}
where $\epsilon_t$ ($= \log\mathcal{E}_t$) denotes the logarithm of the exchange rate.

\subsection{Central bank's loss function} 
\renewcommand{\theequation}{B.\arabic{equation} }
\setcounter{equation}{0}
We now derive the second-order approximation of the household's utility function weighted by degree of openness. The derivation of the central bank's loss function is implemented in the case of policy coordination. The following derivation is based on \citet{clarida2002simple}. 

The utility function of the planner is given by
\begin{align}
W_t = (1-\gamma)\bigg[u \bigg(C_t,\frac{M_t}{P_t}\bigg) -V(N_t)\bigg]+ \gamma\bigg[u\bigg(C^*_t,\frac{M^*_t}{P^*_t}\bigg) -V(N^*_t)\bigg],  \label{eq.util}
\end{align}

In order to obtain a well-defined loss function without calculating the second-order approximation of structural equations, it is necessary to eliminate the distortions caused by monopolistic competition and real money balances. The first distortion is eliminated by an optimal subsidy rate that eliminates the price markup caused by monopolistic competition in each country. At an efficient level,
\begin{align}
\varphi(Y_t^n,Y_t^n,Y_t^{n*}; A_t )= 1, \nonumber
\end{align}
where $\varphi(Y_t^n,Y_t^n,Y_t^{n*}; A_t)$ represents the real marginal cost under efficient output. The fiscal authority chooses the optimal subsidy rate that restores natural output to an efficient level at zero inflation. As mentioned earlier, such an optimal subsidy rate is given by
\begin{align}
(1-\tau)\mu = 1, (1-\tau^*)\mu^* = 1, \nonumber
\end{align}
and, therefore, we obtain $u_c\bar{C}=V_n\bar{N}$. 

The second distortion is a result of an opportunity cost of holding money. As shown in \citet{woodford2003interest}, this opportunity cost should be considerably small in steady state to obtain a well-defined loss function of the central bank. In particular, \citet{woodford2003interest} argues that real money balances are sufficiently close to being satiated in the optimal steady state. To do so, we can eliminate the distortion produced by the opportunity cost of money.\footnote{See Chapter 6 in \citet{woodford2003interest} for a detailed discussion of this issue.}

Before deriving the loss function, we define some notations. First of all, $\bar{H}$ represents the value of steady state, $H_t^n$ is the value of efficient level. Also, we define $h_t= \log(H_t/\bar{H})$ as the deviation of $H_t$ from the steady state. In addition to these notations, we introduce the following equation:
\begin{align}
H_t-\bar{H}=\bar{H}\bigg(\frac{H_t}{\bar{H}}-1 \bigg)\simeq h_t+\frac{1}{2}h_t^2. \nonumber
\end{align}

The second-order approximation of the first term of the right hand side is given by:
\begin{align}
u\bigg(C_t,\frac{M_t}{P_t}\bigg) &\simeq& u_c\bar{C}\bigg[c_t+\frac{1}{2}(1-\sigma)c_t^2 + s_m m_t+\frac{1}{2}(1-\sigma_m) m_t^2 \bigg] +t.i.p+O(\|\xi \|^{3}), \label{eq.aa}
\end{align}
where $t.i.p.$ represents the terms that are independent of monetary policy, and $O(\|\xi \|^{3} )$ indicates that we neglect terms of third or higher order. Also, $m_t = \log(Z_t/\bar{Z})$. In addition, 
\begin{align}
\sigma_m = \frac{u_{mm} \bar{Z}}{u_m}, s_m = -\frac{u_m\bar{Z}}{u_c \bar{C}} \nonumber
\end{align} 
and
\begin{align}
s_m \sigma_m = -(\bar{v}\eta_r)^{-1}, \ \eta_y = \bar{v} \chi \eta_r, \ \chi=\frac{u_{cm}\bar{Z}}{u_m}. \nonumber
\end{align} 
where $\bar{v}$ is the velocity of money and $\bar{Z}=\bar{M}/\bar{P}$.   

Substituting the log-linearization of Eqs. (\ref{eq:29}) and (\ref{eq.hmond}) into Eq. (\ref{eq.aa}), we obtain
\begin{align}
&u\bigg(C_t,\frac{M_t}{P_t}\bigg) \simeq u_c\bar{C}\bigg\{(1-\gamma)y_t+\gamma y_t^* +\frac{1}{2}(1-\sigma)\bigg[(1-\gamma)^2 y_t^2 + \gamma^2 y_t^{*2} + 2(1-\gamma)\gamma y_t y_t^* \bigg] \nonumber \\
&+(1-\gamma)(s_m\eta_y((1-\gamma)y_t+\gamma y_t^*)-\eta_r s_m r_t -\eta_i(\bar{v})^{-1} r_t^2 -\chi^2 \eta_y^2 c_t^2 \bigg\}+t.i.p.+O(\| \xi \|^{3}), \label{eq.bb}
\end{align}

Next, the second-order approximations of the second and third terms of the right side of the utility function are given by:
\begin{align}
&V(N_t) = V_n(\bar{N})\bar(N)\bigg[y_t -a_t + \frac{1}{2}(1+\eta)(y_t -a_t)^2 +\frac{\theta}{2} p_{H,t} \bigg] + t.i.p. + O(||\xi||^3), \label{eq.cc}
\\
&V(N^*_t) = V_n(\bar{N})\bar(N)\bigg[y^*_t -a^*_t + \frac{1}{2}(1+\eta)(y^*_t -a^*_t)^2 +\frac{\theta}{2} p^*_{F,t} \bigg]+ t.i.p. + O(||\xi||^3), \label{eq.dd}
\end{align}
where $p_{H,t}=\int_0^1(P_{H,t}(i)/P_{H,t})^{-\theta}$di and $p^*_{F,t}=\int_0^1(P^*_{F,t}(i)/P^*_{F,t})^{-\theta}$di. 

Combining Eqs.(\ref{eq.bb}), (\ref{eq.cc}), and (\ref{eq.dd}) and using the definition of the natural rate of output for both countries, we obtain
\begin{align}
U_t &\simeq -\frac{u_c\bar{C}}{2}\bigg \{(1-\gamma)[(\sigma+\eta-\gamma(\sigma-1))(y_t-y_t^n)^2+(\bar{v})^{-1}\eta_r r_t^2 +\theta p_{H,t}] \nonumber \\
&\gamma[(\sigma+\eta-(1-\gamma)(\sigma-1))(y^*_t-y^{n*}_t)^2+(\bar{v})^{-1}\eta_r (r^*_t)^2 +\theta p^*_{F,t}] \nonumber \\
&-2\gamma(1-\gamma)(1-\sigma)(y_t-y_t^n)(y^*_t-y^{n*}_t) \bigg \}+t.i.p.+O(\| \xi \|^{3}). \label{eq.ee}
\end{align} 
In this derivation, we used the relationship $u_c \bar{C} = V_n\bar{N}$, which is held in the efficient steady state. Also, following \citet{woodford2003interest}, we assumed that the distortion derived from money holding cost is eliminated in this derivation.

Regarding the term for price dispersion, following \citet{woodford2003interest}, we obtain
\begin{align}
&\sum_{t=0}^{\infty} \beta^t p_{H,t}=\frac{\alpha}{(1-\alpha)(1-\alpha\beta)}\sum_{t=0}^{\infty} \beta^{t} \pi_t^2 +t.i.p.+O(\| \xi \|)^{3}, \label{eq.hh}
\\
&\sum_{t=0}^{\infty} \beta^t p^*_{F,t}=\frac{\alpha^*}{(1-\alpha^*)(1-\alpha^* \beta)}\sum_{t=0}^{\infty} \beta^{t} (\pi_t^*)^2 +t.i.p.+O(\| \xi \|)^{3}. \label{eq.ii}
\end{align}
Substituting Eq. (\ref{eq.hh}) and Eq. (\ref{eq.ii}) into Eq. (\ref{eq.ee}), the central bank's loss function under policy coordination is given by
\begin{align}
\sum_{t=0}^{\infty} W_t \approx -\Omega \sum_{t=0}^{\infty} \beta^t L^w_t + t.i.p. + O(||\xi||^3), \label{eq.welf}
\end{align}
Here, the periodic loss function $L^w_t$ in Equation (\ref{eq.welf}) is given by
\begin{align}
&L^w_t = (1-\psi) \left[\pi_t^2 + \lambda_x x_t^2 + \lambda_r r_t^2\right] + \psi \left[(\pi_t^*)^2 + \lambda_x^* (x_t^*)^2 + \lambda_r^* (r_t^*)^2 \right]-2\Lambda x_t x^*_t, \label{eq.lossfun}
\end{align}
where 
\begin{align}
\varpi = \frac{(1-\alpha)(1-\alpha \beta)}{\alpha}, \ \varpi^* = \frac{(1-\alpha^*)(1-\alpha^* \beta)}{\alpha^*}, \nonumber 
\end{align}
and
\begin{align}
&1-\psi = \frac{(1-\gamma)\varpi^{-1}}{\varpi}, \nonumber \\
&\varpi = (1-\gamma)\varpi^{-1}+\gamma(\varpi^*)^{-1} \nonumber \\
&\lambda_x = \frac{\kappa_1}{\theta}, \ \lambda_r = \frac{\eta_r}{\bar{v}\theta}, \ \lambda_{\Delta \pi}=\frac{(1-\omega)}{\omega\alpha} \nonumber \\
&\lambda^*_x = \frac{\kappa^*_1}{\theta}, \  \lambda^*_r = \frac{\eta_r^*}{\bar{v}^*\theta}, \  \lambda_{\Delta \pi}^*=\frac{(1-\omega^*)}{\omega^*\alpha^*} \nonumber \\
&\Lambda = \frac{2(1-\gamma)\gamma(1-\sigma)}{\varpi\theta} \nonumber
\end{align}

\subsection{Additional quantitative results}
In this section, we report the results of a simulation of the impact of the home country's monetary policy on foreign economies in the case of a global liquidity trap shock when $\sigma<1$. Thus, both countries have a negative international risk-sharing channel. Again, as in our main manuscript, we examine four cases; (a) the foreign country does not face a ZLB constraint, (b) the foreign country faces a ZLB constraint, (c) the foreign country has an FG of the same length as the home country's FG, and (d) the foreign country has an FG of five extra quarters.

The relative risk aversion coefficient is set at 0.5, and the other parameters and the size of the global shock are the same as in our main manuscript. Compared to the case in our main manuscript where the parameter $\sigma$ is greater than one, the first thing to notice from Table C1 is that the decline in welfare in both countries is more significant in all four cases. Second, as the size of the difference between the responses of ZLB and FG policies can be shown in Figures C1-C4, the impact of the home country's monetary policy on the foreign country seems to be smaller when $\sigma<1$ than when $\sigma>1$.

Panel (a) of Table C1 shows that when the home country's central bank adopts FG policy instead of ZLB policy, it contributes to improving the foreign country's welfare in the form of \textit{prosper-thy-neighbor effect}. In this case, if the home country's central bank implements a five-extra-quarters of FG policy, the foreign country's loss (welfare) is minimized (maximized). The other three cases shown in Panels (b), (c), and (d) are similar to Panel (a), where the foreign country's loss (welfare) is minimum (maximum) if the home country's central bank adopts a five-extra-quarters of FG policy.

Thus, even when $\sigma<1$, both central banks can improve their countries' welfare by adopting FG policies when a global liquidity trap occurs. Moreover, since both countries' interests are aligned, the choice of monetary policy will become monotonic. In this case, bargaining and betrayal of determining the length of FG are not beneficial. The coordination of monetary policies is sustained in both countries, unlike the conclusion of our main manuscript.






\setcounter{table}{0}

\renewcommand{\thetable}{\Alph{section}.\arabic{table}}

\begin{table}[H]
\caption{\small{Welfare Losses by a Global Liquidity Trap Shock } }
\begin{footnotesize}
\begin{center} 
(a) Country F without the ZLB constraint 

\begin{tabular}{ccccc} \hline\hline 
H Policy & F Policy & World Losses & H Loss & F Loss  \\ \hline 
ZLB      &  non ZLB   &   43.53   &   20.58   &   22.95   \\ 
FG 2 extra qrts   &  non ZLB  & 40.96  &    18.24  &    22.72   \\ 
FG 4 extra qrts  &  non ZLB & 35.42   &   12.91   &   22.51  \\ 
FG 5 extra qrts &  non ZLB &  \bf{33.84}   & \bf{11.28} & \bf{22.55}     \\
FG 6 extra qrts &  non ZLB &  34.96  &  12.23   &   22.73   \\  \hline 
\end{tabular}%
\end{center}

\begin{center} 
(b) Country F with the ZLB constraint 

\begin{tabular}{ccccc} \hline\hline 
 H Policy & F Policy & World Losses & H Loss & F Loss  \\ \hline 
ZLB      &   ZLB   & 47.12   &   23.56    &  23.56   \\ 
FG 2 extra qrts  &  ZLB  & 44.08   &   20.89  &    23.20   \\ 
FG 4 extra qrts  &  ZLB &  37.70  &    14.75   & 22.95 \\ 
FG 5 extra qrts &  ZLB &   \bf{35.62}  &   \bf{12.57}   &   \bf{23.05}    \\
FG 6 extra qrts &  ZLB &  36.20 & 12.87   &   23.32   \\  \hline
\end{tabular}%
\end{center}

\begin{center} 
(c) Countries H and F adopt the Same Length of FG 
\begin{tabular}{ccccc} \hline\hline 
H Policy & F Policy & World Losses & H Loss & F Loss  \\ \hline 
ZLB      &   ZLB   & 47.12   &   23.56   &   23.56   \\ 
FG 2 extra qrts &  FG 2 extra qrts  &  39.75   &   19.87   &   19.87   \\ 
FG 4 extra qrts  &  FG 4 extra qrts &  25.11   &   12.56   &   12.56  \\ 
FG 5 extra qrts &  FG 5 extra qrts & \bf{23.29 }   &   \bf{11.65}   &   \bf{11.65}   \\
FG 6 extra qrts &  FG 6 extra qrts & 32.05  &    16.03   &   16.03  \\
 \hline
\end{tabular}%
\end{center}

\begin{center} 
(d) Country F takes fixed 6 extra qrts FG  

\begin{tabular}{ccccc} \hline\hline 
H Policy & F Policy & World Losses & H Loss & F Loss  \\ \hline 
ZLB      &  FG 5 extra qrts   &  35.62    &  23.05  &    12.57  \\ 
FG 2 extra qrts  &  FG 5 extra qrts  & 31.04  &    18.87  &    12.17  \\ 
FG 4 extra qrts  &  FG 5 extra qrts & {23.84}   &   {12.17}   &  {11.67}    \\ 
FG 5 extra qrts &  FG 5 extra qrts & \bf{23.29 }   &   \bf{11.65}   &   \bf{11.65}   \\
FG 6 extra qrts &  FG 5 extra qrts &  26.30   &   14.63   &   11.67   \\ \hline
\end{tabular}%
\end{center}
\end{footnotesize}
\end{table}

\newpage
\renewcommand{\thefigure}{\Alph{section}.\arabic{figure}}
\setcounter{figure}{0}
\begin{figure}[H]
\caption{The impulse response to the global liquidity trap shock: Country F without the ZLB}
\includegraphics[width=13cm,height=8cm]{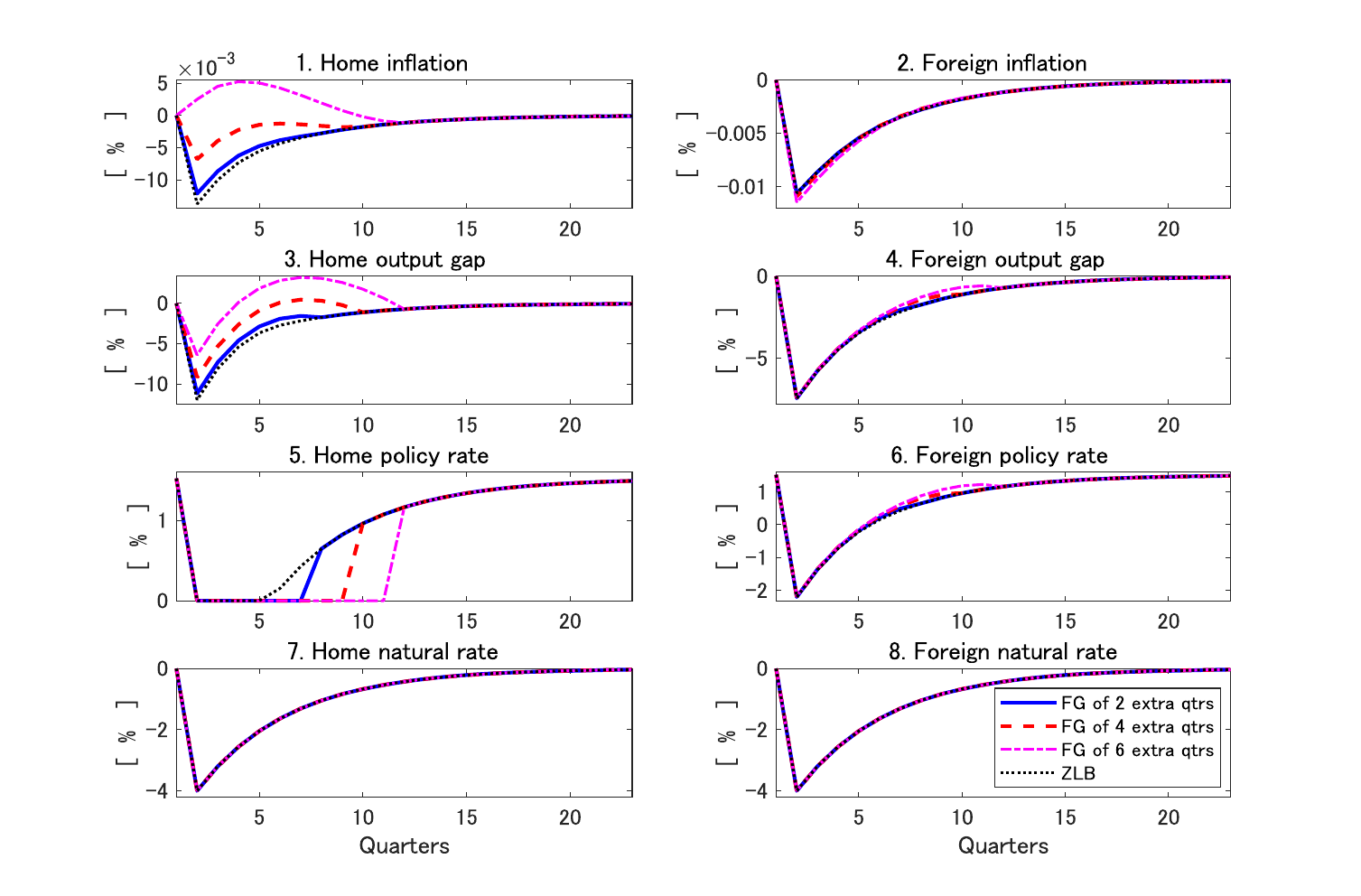}
\end{figure}
\begin{figure}[H]
\caption{The impulse response to the global liquidity trap: Country F with the ZLB}
\includegraphics[width=13cm,height=8cm]{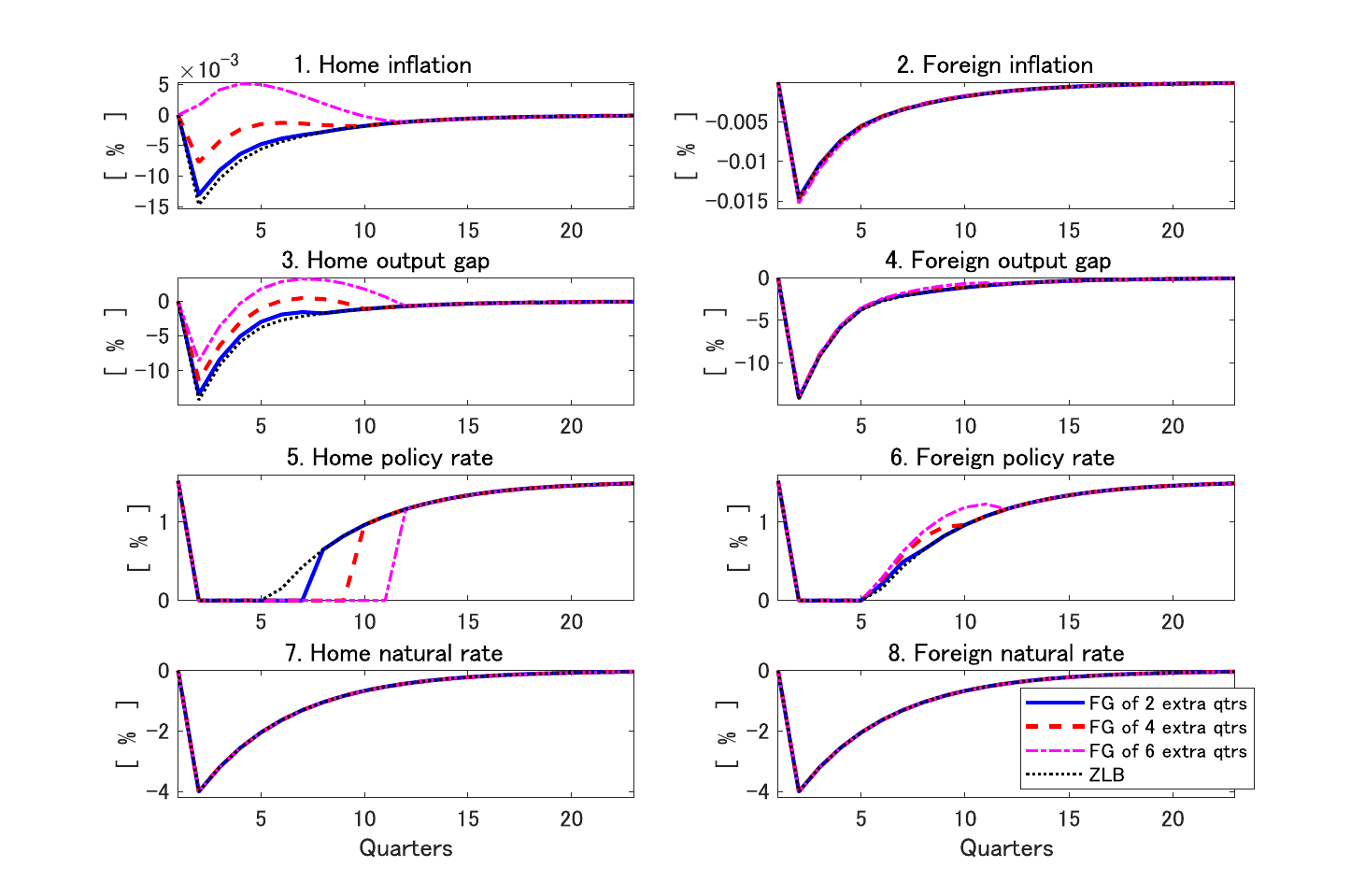}
\end{figure}

\begin{figure}[H]
\caption{The impulse response to the global liquidity trap shock: Country H and F adopt the same length of FG}
\includegraphics[width=13cm,height=8cm]{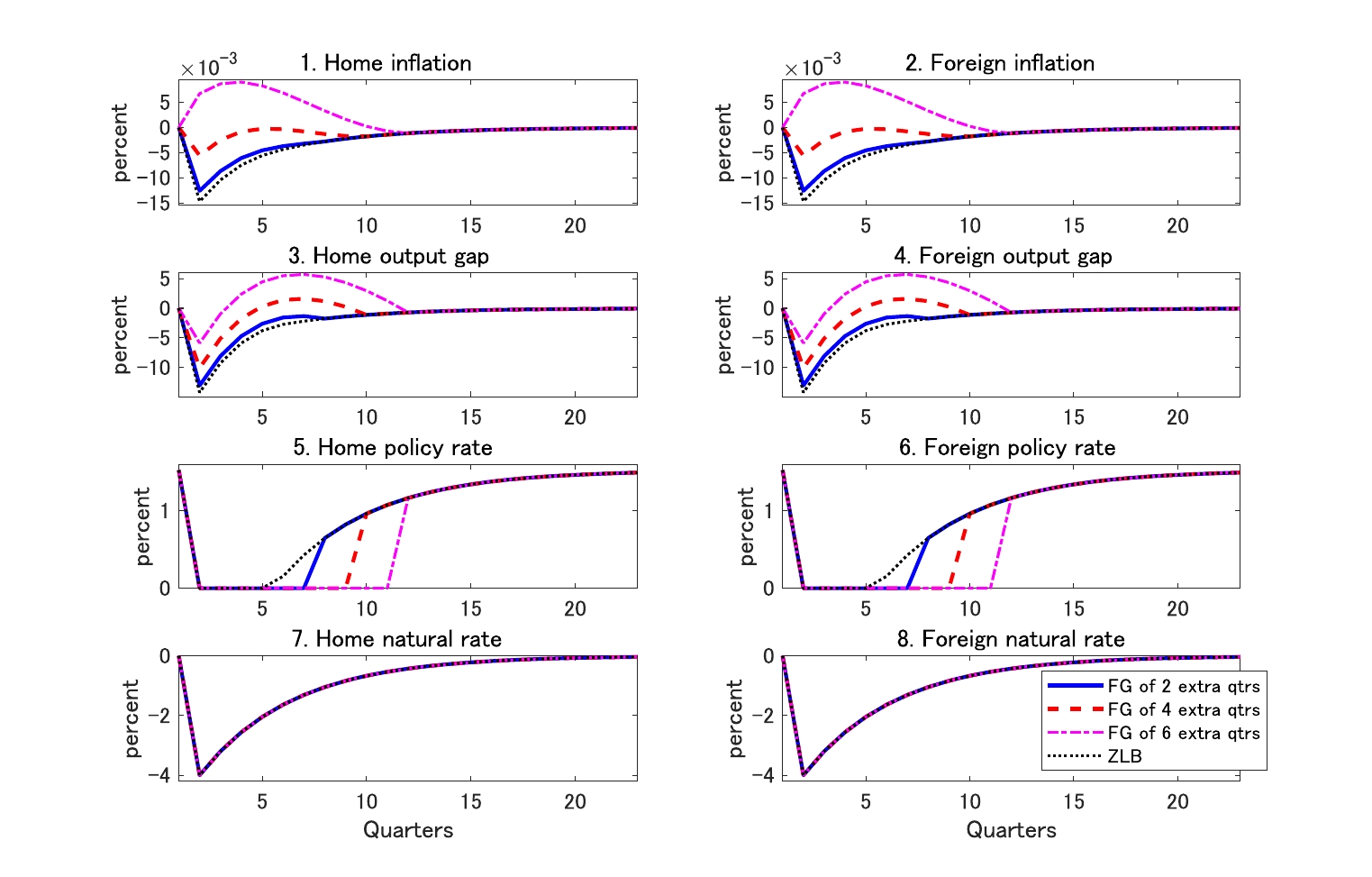}
\end{figure}

\begin{figure}[H]
\caption{The impulse response to the global liquidity trap shock: Country F takes six-period fixed-length FG}
\includegraphics[width=13cm,height=8cm]{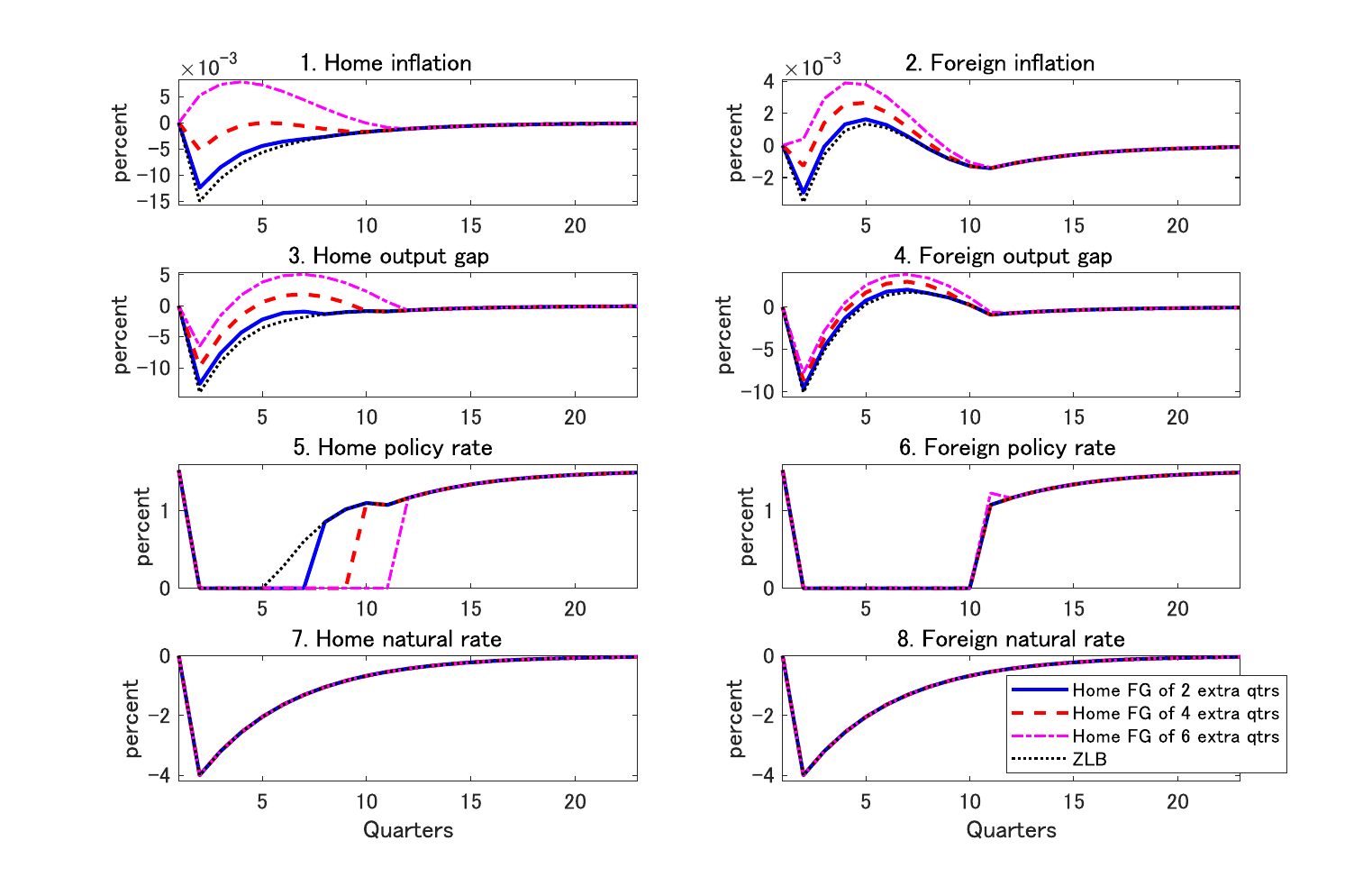}
\end{figure}


\end{appendix}

%% file: 20240521_main.bbl
\begin{thebibliography}{44}
\expandafter\ifx\csname natexlab\endcsname\relax\def\natexlab#1{#1}\fi
\providecommand{\url}[1]{\texttt{#1}}
\providecommand{\href}[2]{#2}
\providecommand{\path}[1]{#1}
\providecommand{\DOIprefix}{doi:}
\providecommand{\ArXivprefix}{arXiv:}
\providecommand{\URLprefix}{URL: }
\providecommand{\Pubmedprefix}{pmid:}
\providecommand{\doi}[1]{\href{http://dx.doi.org/#1}{\path{#1}}}
\providecommand{\Pubmed}[1]{\href{pmid:#1}{\path{#1}}}
\providecommand{\bibinfo}[2]{#2}
\ifx\xfnm\relax \def\xfnm[#1]{\unskip,\space#1}\fi
\bibitem[{Adam and Billi(2006)}]{adam2006optimal}
\bibinfo{author}{Adam, K.}, \bibinfo{author}{Billi, R.M.}, \bibinfo{year}{2006}.
\newblock \bibinfo{title}{Optimal monetary policy under commitment with a zero bound on nominal interest rates}.
\newblock \bibinfo{journal}{Journal of Money, Credit and Banking} , \bibinfo{pages}{1877--1905}.
\bibitem[{Adam and Billi(2007)}]{adam2007discretionary}
\bibinfo{author}{Adam, K.}, \bibinfo{author}{Billi, R.M.}, \bibinfo{year}{2007}.
\newblock \bibinfo{title}{Discretionary monetary policy and the zero lower bound on nominal interest rates}.
\newblock \bibinfo{journal}{Journal of Monetary Economics} \bibinfo{volume}{54}, \bibinfo{pages}{728--752}.
\bibitem[{Alpanda and Kabaca(2020)}]{alpanda2020international}
\bibinfo{author}{Alpanda, S.}, \bibinfo{author}{Kabaca, S.}, \bibinfo{year}{2020}.
\newblock \bibinfo{title}{International spillovers of large-scale asset purchases}.
\newblock \bibinfo{journal}{Journal of the European Economic Association} \bibinfo{volume}{18}, \bibinfo{pages}{342--391}.
\bibitem[{Benigno and L\'opez-Salido(2006)}]{benigno2006jmcb}
\bibinfo{author}{Benigno, P.}, \bibinfo{author}{L\'opez-Salido, D.J.}, \bibinfo{year}{2006}.
\newblock \bibinfo{title}{{Inflation persistence and optimal monetary policy in the Euro area}}.
\newblock \bibinfo{journal}{Journal of Money, Credit, and Banking} \bibinfo{volume}{38}, \bibinfo{pages}{587--614}.
\bibitem[{Bernanke and Reinhart(2004)}]{bernanke2004conducting}
\bibinfo{author}{Bernanke, B.S.}, \bibinfo{author}{Reinhart, V.R.}, \bibinfo{year}{2004}.
\newblock \bibinfo{title}{Conducting monetary policy at very low short-term interest rates}.
\newblock \bibinfo{journal}{American Economic Review} \bibinfo{volume}{94}, \bibinfo{pages}{85--90}.
\bibitem[{Bhattarai et~al.(2021)Bhattarai, Chatterjee and Park}]{bhattarai2021effects}
\bibinfo{author}{Bhattarai, S.}, \bibinfo{author}{Chatterjee, A.}, \bibinfo{author}{Park, W.Y.}, \bibinfo{year}{2021}.
\newblock \bibinfo{title}{{Effects of US quantitative easing on emerging market economies}}.
\newblock \bibinfo{journal}{Journal of Economic Dynamics and Control} \bibinfo{volume}{122}, \bibinfo{pages}{104031}.
\bibitem[{Bodenstein et~al.(2012)Bodenstein, Hebden and Nunes}]{bodenstein2012imperfect}
\bibinfo{author}{Bodenstein, M.}, \bibinfo{author}{Hebden, J.}, \bibinfo{author}{Nunes, R.}, \bibinfo{year}{2012}.
\newblock \bibinfo{title}{Imperfect credibility and the zero lower bound}.
\newblock \bibinfo{journal}{Journal of Monetary Economics} \bibinfo{volume}{59}, \bibinfo{pages}{135--149}.
\bibitem[{Boehl and Strobel(2024)}]{boehl2024estimation}
\bibinfo{author}{Boehl, G.}, \bibinfo{author}{Strobel, F.}, \bibinfo{year}{2024}.
\newblock \bibinfo{title}{Estimation of dsge models with the effective lower bound}.
\newblock \bibinfo{journal}{Journal of Economic Dynamics and Control} \bibinfo{volume}{158}, \bibinfo{pages}{104784}.
\bibitem[{Boneva et~al.(2018)Boneva, Harrison and Waldron}]{boneva2018threshold}
\bibinfo{author}{Boneva, L.}, \bibinfo{author}{Harrison, R.}, \bibinfo{author}{Waldron, M.}, \bibinfo{year}{2018}.
\newblock \bibinfo{title}{Threshold-based forward guidance}.
\newblock \bibinfo{journal}{Journal of Economic Dynamics and Control} \bibinfo{volume}{90}, \bibinfo{pages}{138--155}.
\bibitem[{Calvo(1983)}]{calvo1983staggered}
\bibinfo{author}{Calvo, G.A.}, \bibinfo{year}{1983}.
\newblock \bibinfo{title}{Staggered prices in a utility-maximizing framework}.
\newblock \bibinfo{journal}{Journal of Monetary Economics} \bibinfo{volume}{12}, \bibinfo{pages}{383--398}.
\bibitem[{Campbell et~al.(2019)Campbell, Ferroni, Fisher and Melosi}]{campbell2019limits}
\bibinfo{author}{Campbell, J.R.}, \bibinfo{author}{Ferroni, F.}, \bibinfo{author}{Fisher, J.D.}, \bibinfo{author}{Melosi, L.}, \bibinfo{year}{2019}.
\newblock \bibinfo{title}{The limits of forward guidance}.
\newblock \bibinfo{journal}{Journal of Monetary Economics} \bibinfo{volume}{108}, \bibinfo{pages}{118--134}.
\bibitem[{Carlstrom et~al.(2015)Carlstrom, Fuerst and Paustian}]{carlstrom2015inflation}
\bibinfo{author}{Carlstrom, C.T.}, \bibinfo{author}{Fuerst, T.S.}, \bibinfo{author}{Paustian, M.}, \bibinfo{year}{2015}.
\newblock \bibinfo{title}{{Inflation and output in New Keynesian models with a transient interest rate peg}}.
\newblock \bibinfo{journal}{Journal of Monetary Economics} \bibinfo{volume}{76}, \bibinfo{pages}{230--243}.
\bibitem[{Clarida et~al.(2002)Clarida, Gal{\i} and Gertler}]{clarida2002simple}
\bibinfo{author}{Clarida, R.}, \bibinfo{author}{Gal{\i}, J.}, \bibinfo{author}{Gertler, M.}, \bibinfo{year}{2002}.
\newblock \bibinfo{title}{A simple framework for international monetary policy analysis}.
\newblock \bibinfo{journal}{Journal of Monetary Economics} \bibinfo{volume}{49}, \bibinfo{pages}{879--904}.
\bibitem[{Cole and Martínez-García(2023)}]{cole_2023}
\bibinfo{author}{Cole, S.J.}, \bibinfo{author}{Martínez-García, E.}, \bibinfo{year}{2023}.
\newblock \bibinfo{title}{{The effect of central bank credibility on forward guidance in an estimated New Keynesian model}}.
\newblock \bibinfo{journal}{Macroeconomic Dynamics} \bibinfo{volume}{27}, \bibinfo{pages}{532–570}.
\bibitem[{Cook and Devereux(2011)}]{cook2011optimal}
\bibinfo{author}{Cook, D.}, \bibinfo{author}{Devereux, M.B.}, \bibinfo{year}{2011}.
\newblock \bibinfo{title}{Optimal fiscal policy in a world liquidity trap}.
\newblock \bibinfo{journal}{European Economic Review} \bibinfo{volume}{55}, \bibinfo{pages}{443--462}.
\bibitem[{Del~Negro et~al.(2012)Del~Negro, Giannoni and Patterson}]{del2012forward}
\bibinfo{author}{Del~Negro, M.}, \bibinfo{author}{Giannoni, M.P.}, \bibinfo{author}{Patterson, C.}, \bibinfo{year}{2012}.
\newblock \bibinfo{title}{The forward guidance puzzle}.
\newblock \bibinfo{journal}{FRB of New York Staff Report} .
\bibitem[{Eggertsson and Woodford(2003)}]{woodford2003zero}
\bibinfo{author}{Eggertsson, G.}, \bibinfo{author}{Woodford, M.}, \bibinfo{year}{2003}.
\newblock \bibinfo{title}{The zero bound on interest rates and optimal monetary policy}.
\newblock \bibinfo{journal}{Brookings Papers on Economic Activity} \bibinfo{volume}{1}, \bibinfo{pages}{139--233}.
\bibitem[{Eggertsson et~al.(2020)Eggertsson, Egiev, Lin, Platzer and Riva}]{eggertsson2020toolkit}
\bibinfo{author}{Eggertsson, G.B.}, \bibinfo{author}{Egiev, S.K.}, \bibinfo{author}{Lin, A.}, \bibinfo{author}{Platzer, J.}, \bibinfo{author}{Riva, L.}, \bibinfo{year}{2020}.
\newblock \bibinfo{title}{A Toolkit for Solving Models with a Lower Bound on Interest Rates of Stochastic Duration}.
\newblock \bibinfo{type}{Technical Report}. National Bureau of Economic Research.
\bibitem[{English et~al.(2013)English, L{\'o}pez-Salido and Tetlow}]{englishfederal}
\bibinfo{author}{English, W.B.}, \bibinfo{author}{L{\'o}pez-Salido, J.D.}, \bibinfo{author}{Tetlow, R.J.}, \bibinfo{year}{2013}.
\newblock \bibinfo{title}{The federal reserve's framework for monetary policy—recent changes and new questions} .
\bibitem[{Fujiwara et~al.(2013)Fujiwara, Nakajima, Sudo and Teranishi}]{fujiwara2013global}
\bibinfo{author}{Fujiwara, I.}, \bibinfo{author}{Nakajima, T.}, \bibinfo{author}{Sudo, N.}, \bibinfo{author}{Teranishi, Y.}, \bibinfo{year}{2013}.
\newblock \bibinfo{title}{Global liquidity trap}.
\newblock \bibinfo{journal}{Journal of Monetary Economics} \bibinfo{volume}{60}, \bibinfo{pages}{936--949}.
\bibitem[{Gabaix(2020)}]{gabaix2020behavioral}
\bibinfo{author}{Gabaix, X.}, \bibinfo{year}{2020}.
\newblock \bibinfo{title}{{A behavioral New Keynesian model}}.
\newblock \bibinfo{journal}{American Economic Review} \bibinfo{volume}{110}, \bibinfo{pages}{2271--2327}.
\bibitem[{Gal{\'\i}(2020)}]{gali2020uncovered}
\bibinfo{author}{Gal{\'\i}, J.}, \bibinfo{year}{2020}.
\newblock \bibinfo{title}{Uncovered Interest Parity, Forward Guidance, and the Exchange Rate}.
\newblock \bibinfo{type}{Technical Report}. National Bureau of Economic Research.
\bibitem[{Guerrieri and Iacoviello(2015)}]{guerrieri2015occbin}
\bibinfo{author}{Guerrieri, L.}, \bibinfo{author}{Iacoviello, M.}, \bibinfo{year}{2015}.
\newblock \bibinfo{title}{Occbin: A toolkit for solving dynamic models with occasionally binding constraints easily}.
\newblock \bibinfo{journal}{Journal of Monetary Economics} \bibinfo{volume}{70}, \bibinfo{pages}{22--38}.
\bibitem[{Haberis et~al.(2019)Haberis, Harrison and Waldron}]{haberis2019uncertain}
\bibinfo{author}{Haberis, A.}, \bibinfo{author}{Harrison, R.}, \bibinfo{author}{Waldron, M.}, \bibinfo{year}{2019}.
\newblock \bibinfo{title}{Uncertain policy promises}.
\newblock \bibinfo{journal}{European Economic Review} \bibinfo{volume}{111}, \bibinfo{pages}{459--474}.
\bibitem[{Haberis and Lipi{\'n}ska(2020)}]{haberis2020welfare}
\bibinfo{author}{Haberis, A.}, \bibinfo{author}{Lipi{\'n}ska, A.}, \bibinfo{year}{2020}.
\newblock \bibinfo{title}{A welfare-based analysis of international monetary policy spillovers at the zero lower bound}.
\newblock \bibinfo{journal}{Journal of Money, Credit and Banking} \bibinfo{volume}{52}, \bibinfo{pages}{1107--1145}.
\bibitem[{Hirose et~al.(2023)Hirose, Iiboshi, Shintani and Ueda}]{hirose2023estimating}
\bibinfo{author}{Hirose, Y.}, \bibinfo{author}{Iiboshi, H.}, \bibinfo{author}{Shintani, M.}, \bibinfo{author}{Ueda, K.}, \bibinfo{year}{2023}.
\newblock \bibinfo{title}{{Estimating a behavioral new Keynesian model with the zero lower bound}}.
\newblock \bibinfo{journal}{Journal of Money, Credit, and Banking} .
\bibitem[{Ida(2013)}]{ida2013optimal}
\bibinfo{author}{Ida, D.}, \bibinfo{year}{2013}.
\newblock \bibinfo{title}{Optimal monetary policy rules in a two-country economy with a zero bound on nominal interest rates}.
\newblock \bibinfo{journal}{The North American Journal of Economics and Finance} \bibinfo{volume}{24}, \bibinfo{pages}{223--242}.
\bibitem[{Ida(2023)}]{ida2023effect}
\bibinfo{author}{Ida, D.}, \bibinfo{year}{2023}.
\newblock \bibinfo{title}{The effect of real money balances on international monetary policy transmission}.
\newblock \bibinfo{journal}{Journal of International Money and Finance} \bibinfo{volume}{139}, \bibinfo{pages}{102964}.
\bibitem[{Iiboshi et~al.(2022)Iiboshi, Shintani and Ueda}]{iiboshi2022estimating}
\bibinfo{author}{Iiboshi, H.}, \bibinfo{author}{Shintani, M.}, \bibinfo{author}{Ueda, K.}, \bibinfo{year}{2022}.
\newblock \bibinfo{title}{{Estimating a nonlinear new Keynesian model with the zero lower bound for Japan}}.
\newblock \bibinfo{journal}{Journal of Money, Credit and Banking} \bibinfo{volume}{54}, \bibinfo{pages}{1637--1671}.
\bibitem[{Jones et~al.(2018)Jones, Kulish and Rees}]{jones2018international}
\bibinfo{author}{Jones, C.}, \bibinfo{author}{Kulish, M.}, \bibinfo{author}{Rees, D.M.}, \bibinfo{year}{2018}.
\newblock \bibinfo{title}{International Spillovers of Forward Guidance Shocks}.
\newblock \bibinfo{publisher}{International Monetary Fund}.
\bibitem[{Jung et~al.(2005)Jung, Teranishi and Watanabe}]{jung2005optimal}
\bibinfo{author}{Jung, T.}, \bibinfo{author}{Teranishi, Y.}, \bibinfo{author}{Watanabe, T.}, \bibinfo{year}{2005}.
\newblock \bibinfo{title}{Optimal monetary policy at the zero-interest-rate bound}.
\newblock \bibinfo{journal}{Journal of Money, Credit, and Banking} \bibinfo{volume}{37}, \bibinfo{pages}{813--835}.
\bibitem[{Kolasa and Weso{\l}owski(2020)}]{kolasa2020international}
\bibinfo{author}{Kolasa, M.}, \bibinfo{author}{Weso{\l}owski, G.}, \bibinfo{year}{2020}.
\newblock \bibinfo{title}{International spillovers of quantitative easing}.
\newblock \bibinfo{journal}{Journal of International Economics} \bibinfo{volume}{126}, \bibinfo{pages}{103330}.
\bibitem[{McKay et~al.(2016)McKay, Nakamura and Steinsson}]{mckay2016power}
\bibinfo{author}{McKay, A.}, \bibinfo{author}{Nakamura, E.}, \bibinfo{author}{Steinsson, J.}, \bibinfo{year}{2016}.
\newblock \bibinfo{title}{The power of forward guidance revisited}.
\newblock \bibinfo{journal}{American Economic Review} \bibinfo{volume}{106}, \bibinfo{pages}{3133--58}.
\bibitem[{Moessner and Rungcharoenkitkul(2019)}]{moessner2019zero}
\bibinfo{author}{Moessner, R.}, \bibinfo{author}{Rungcharoenkitkul, P.}, \bibinfo{year}{2019}.
\newblock \bibinfo{title}{The zero lower bound, forward guidance and how markets respond to news}.
\newblock \bibinfo{journal}{BIS Quarterly Review, March} .
\bibitem[{Nakajima(2008)}]{nakajima2008liquidity}
\bibinfo{author}{Nakajima, T.}, \bibinfo{year}{2008}.
\newblock \bibinfo{title}{Liquidity trap and optimal monetary policy in open economies}.
\newblock \bibinfo{journal}{Journal of the Japanese and international Economies} \bibinfo{volume}{22}, \bibinfo{pages}{1--33}.
\bibitem[{Nakata et~al.(2019)Nakata, Ogaki, Schmidt and Yoo}]{nakata2019attenuating}
\bibinfo{author}{Nakata, T.}, \bibinfo{author}{Ogaki, R.}, \bibinfo{author}{Schmidt, S.}, \bibinfo{author}{Yoo, P.}, \bibinfo{year}{2019}.
\newblock \bibinfo{title}{Attenuating the forward guidance puzzle: implications for optimal monetary policy}.
\newblock \bibinfo{journal}{Journal of Economic Dynamics and Control} \bibinfo{volume}{105}, \bibinfo{pages}{90--106}.
\bibitem[{Obstfeld and Rogoff(1996)}]{obstfeld1996foundations}
\bibinfo{author}{Obstfeld, M.}, \bibinfo{author}{Rogoff, K.}, \bibinfo{year}{1996}.
\newblock \bibinfo{title}{{Foundations of International Macroeconomics}}.
\newblock \bibinfo{publisher}{MIT press}.
\bibitem[{Pappa(2004)}]{pappa2004ecb}
\bibinfo{author}{Pappa, E.}, \bibinfo{year}{2004}.
\newblock \bibinfo{title}{{Do the ECB and the Fed really need to cooperate? Optimal monetary policy in a two-country world}}.
\newblock \bibinfo{journal}{Journal of Monetary Economics} \bibinfo{volume}{51}, \bibinfo{pages}{753--779}.
\bibitem[{Taylor(1993)}]{taylor1993discretion}
\bibinfo{author}{Taylor, J.B.}, \bibinfo{year}{1993}.
\newblock \bibinfo{title}{Discretion versus policy rules in practice}.
\newblock \bibinfo{journal}{Carnegie-Rochester conference series on public policy} \bibinfo{volume}{39}, \bibinfo{pages}{195--214}.
\bibitem[{Tille(2001)}]{tille2001role}
\bibinfo{author}{Tille, C.}, \bibinfo{year}{2001}.
\newblock \bibinfo{title}{The role of consumption substitutability in the international transmission of monetary shocks}.
\newblock \bibinfo{journal}{Journal of International Economics} \bibinfo{volume}{53}, \bibinfo{pages}{421--444}.
\bibitem[{Walsh(2017)}]{walsh2017monetary}
\bibinfo{author}{Walsh, C.E.}, \bibinfo{year}{2017}.
\newblock \bibinfo{title}{{Monetary Theory and Policy}}.
\newblock \bibinfo{publisher}{MIT press}.
\bibitem[{Williamson et~al.(2015)}]{williamson2015monetary}
\bibinfo{author}{Williamson, S.D.}, et~al., \bibinfo{year}{2015}.
\newblock \bibinfo{title}{Monetary policy normalization in the united states}.
\newblock \bibinfo{journal}{Federal Reserve Bank of St. Louis Review} \bibinfo{volume}{97}, \bibinfo{pages}{87--108}.
\bibitem[{Woodford(2003)}]{woodford2003interest}
\bibinfo{author}{Woodford, M.}, \bibinfo{year}{2003}.
\newblock \bibinfo{title}{{Interest and Prices: Foundations of a Theory of Monetary Policy}}.
\newblock \bibinfo{publisher}{Princeton University Press}.
\bibitem[{Wu and Zhang(2019)}]{wu2019global}
\bibinfo{author}{Wu, J.C.}, \bibinfo{author}{Zhang, J.}, \bibinfo{year}{2019}.
\newblock \bibinfo{title}{Global effective lower bound and unconventional monetary policy}.
\newblock \bibinfo{journal}{Journal of International Economics} \bibinfo{volume}{118}, \bibinfo{pages}{200--216}.

\end{thebibliography}
